\def\tprod{\mathop{\textstyle \prod }}
\def\tsum{\mathop{\textstyle \sum }}
\def\dsum{\mathop{\displaystyle \sum }}
\def\func#1{\mathop{\rm #1}\nolimits}

\documentclass[12pt]{article}

\usepackage{amsmath}
\usepackage{amssymb}

\usepackage{fancyhdr}
\renewcommand{\maketitle}{
    \begin{center}
      \Large
        {\bf Harmonic Oscillator States \\ with Non-Integer Orbital Angular Momentum}
        \vskip .3 true cm
      \normalsize
        Martin Land \\
        Department of Computer Science \\
        Hadassah College, Jerusalem \\ 
        email: martin@hadassah.ac.il
      \end{center}
      \vskip .5 true cm
}
\begin{document}

\title{}
\author{}
\maketitle

\begin{abstract}
We study the quantum mechanical harmonic oscillator in two and three
dimensions, with particular attention to the solutions as basis states for
representing their respective symmetry groups --- O(2), O(3), and O(2,1).
Solving the Schrodinger equation by separating variables in polar
coordinates, we obtain wavefunctions characterized by a principal quantum
number, the group Casimir eigenvalue, and one observable component of
orbital angular momentum, with eigenvalue $m+s$, for integer $m$ and real
constant parameter $s$. For each of the three symmetry groups, $s$ splits
the solutions into two inequivalent representations, one associated with $s=0
$, from which we recover the familiar description of the oscillator as a
product of one-dimensional solutions, and the other with $s>0$ (in three
dimensions, solutions are found for $s=0$ and $s=1/2$) whose solutions are
non-separable in Cartesian coodinates, and are hence overlooked by the
standard Fock space approach. In two dimensions, a single set of creation
and annihilation operators forms a ladder representation for the allowed
oscillator states for any $s$, and the degeneracy of energy states is always
finite. However, in three dimensions, the integer and half-integer
eigenstates are qualitatively different: the former can be expressed as
finite dimensional irreducible tensors under O(3) or O(2,1) while the latter
exhibit infinite degeneracy. Creation operators that produce the allowed
integer states by acting on the non-degenerate ground state are constructed
as irreducible tensor products of the fundamental vector representation.
However, since the half-integer ground state has infinite degeneracy, the
vector representation of the creation operators does not take this ground
state to the calcluated first excited level, and the general construction
does not act as a ladder representation for the half-integer states. For all 
$s\neq 0$ solutions, the SU(N) symmetry of the harmonic oscillator
Hamiltonian recently discussed by Bars is spontaneously broken by the ground
state. The connection of this symmetry breaking to the non-separability into
one-dimensional Cartesian solutions is demonstrated.
\end{abstract}

\baselineskip7mm \parindent=0cm \parskip=10pt

\section{Introduction}

Along with its classical counterpart, the quantum harmonic oscillator is a
well-studied model with exact solutions and connections to many physical
systems for which it serves as foundation or approximation. Beyond its
application to atomic and molecular spectra, statistical mechanics, and by
way of various relativistic generalizations to quark dynamics, certain
general techniques associated with the harmonic oscillator, including the
Fock space ensemble of uncoupled modes and Dirac's factorization of the
Hamiltonian into creation and annihilation operators, serve as conceptual
building blocks in areas ranging from blackbody radiation to canonical
quantization and string theory. Yet, despite the subject's long history,
fundamental new insights continue to emerge \cite{Kastrup, Bars}.

In this paper, we study the quantum mechanical harmonic oscillator in two
and three dimensions, with emphasis on the solutions as basis states for
representations of their respective symmetry groups --- O(2), O(3), and
O(2,1). The original motivation for this work was an attempt to develop a
ladder representation of creation and annihilation operators for the
relativistic oscillator model found by Horwitz and Arshansky \cite{bound}
who applied a covariant formulation of quantum mechanics \cite{H-P} to
relativistic generalizations of the classical central force bound state
problems. These models, which are obtained by inducing a representation of
O(3,1) on wavefunctions whose dynamics are restricted to the spacelike
sector of an O(2,1)-invariant subspace, exhibit a positive spectrum, and
belong to half-integral representations of O(3,1). According to a virial
theorem \cite{talk} for the covariant quantum mechanics, the restriction to
spacelike dynamics guarantees a positive spectrum, but since there is no
obvious way to realize this nonholonomic constraint in Cartesian
coordinates, the eigenvalue equation was posed in a hyperspherical
parameterization. To address the unusual characteristics of these solutions,
we sought to develop a creation/annihilation algebra associated with polar
coordinates and non-integer orbital angular momentum. Although the algebraic
approach succeeds in reproducing the basic oscillator features for both
integer and non-integer representations in two dimensions and for the
integer representations in three dimensions, Dirac's factorization of the
Hamiltonian does not lead to creation/annihilation operators for the
half-integer representations of O(3) or O(2,1). This paper presents a
summary of results to be demonstrated in greater detail in a subsequent
paper. To expose the common features of the oscillators associated with the
three symmetries, we develop the models in tandem and use common notation,
as far as is possible.

We write the harmonic oscillator Hamiltonian
\begin{equation}
H=\frac{1}{2}\left( p^{2}+\omega ^{2}x^{2}\right) =\frac{1}{2}\eta _{\mu \nu
}\left( p^{\mu }p^{\nu }+\omega ^{2}x^{\mu }x^{\nu }\right) 
\end{equation}%
to describe either an O($D$) nonrelativistic oscillator with Euclidean metric%
\begin{equation}
\eta _{\mu \nu }=\delta _{\mu \nu },\;\;\mu ,\nu =1,...,D
\end{equation}%
or an O($D-1,1$) relativistic oscillator with Lorentz metric%
\begin{equation}
\eta _{\mu \nu }=\text{diag}\left( -1,1,...,1\right) ,\;\;\mu ,\nu
=0,...,D-1.
\end{equation}%
The standard approach to Fock space proceeds by separation of Cartesian
variables and subsequent application of Dirac's factorization of the
one-dimensional Hamiltonian for each degree of freedom. Assuming a product
solution of one-dimensional oscillators 
\begin{equation}
\psi \left( x\right) =\tprod\limits_{\mu }\psi \left( x^{\mu }\right) %
\mbox{\qquad\qquad}E=\tsum_{\mu }E^{\mu }
\end{equation}%
the Hamiltonian separates into a sum of $D$ mode-number terms as%
\begin{equation}
H=\omega \eta _{\mu \nu }\left( \bar{a}^{\mu }a^{\nu }+\frac{1}{2}\eta ^{\mu
\nu }\right) =\omega \tsum_{\mu }\eta _{\mu \mu }\left( N^{\mu }+\frac{1}{2}%
\eta ^{\mu \mu }\right)   \label{hamil}
\end{equation}%
with creation/annihilation operators%
\begin{equation}
a^{\mu }=\frac{1}{\sqrt{2}}\left( x^{\mu }+ip^{\mu }\right) %
\mbox{\qquad\qquad}\overline{a}^{\mu }=\frac{1}{\sqrt{2}}\left( x^{\mu
}-ip^{\mu }\right)  \label{cr/an}
\end{equation}%
that satisfy%
\begin{equation}
\left[ a^{\mu },\overline{a}^{\nu }\right] =\eta ^{\mu \nu }
\end{equation}%
and mode number operators $N^{\mu }=\bar{a}^{\mu }a^{\mu }$ (no summation)
that satisfy%
\begin{equation}
\left[ N^{\mu },\overline{a}^{\nu }\right] =\eta ^{\mu \nu }\overline{a}%
^{\mu }\mbox{\qquad}\left[ N^{\mu },a^{\nu }\right] =-\eta ^{\mu \nu }a^{\mu
}\mbox{\qquad}\left[ N^{\mu },N^{\nu }\right] =0.
\end{equation}%
The products $\left\vert n\right\rangle =\tprod\limits_{\mu }\left\vert
n^{\mu }\right\rangle $ of $N^{\mu }$ eigenstates form a Fock space of
orthogonal oscillator modes with the ladder property%
\begin{equation}
\bar{a}^{\mu }\left\vert n\right\rangle =e^{i\phi _{+}^{\mu }}\sqrt{n^{\mu
}+\eta ^{\mu \mu }}\left\vert n+\eta ^{\mu \mu }\mathbf{e}_{\mu
}\right\rangle \mbox{\qquad}a^{\mu }\left\vert n\right\rangle =e^{i\phi
_{-}^{\mu }}\sqrt{n^{\mu }}\left\vert n-\eta ^{\mu \mu }\mathbf{e}_{\mu
}\right\rangle ,  \label{ladder-1}
\end{equation}%
where the $\mathbf{e}_{\mu }$ are unit vectors in the occupation number
space 
\begin{equation}
\left( \mathbf{e}_{\mu }\right) ^{\lambda }=\delta _{\mu }^{\lambda }
\end{equation}%
and the particular choice of phases $e^{i\phi _{+}^{\mu }}$ and $e^{i\phi
_{-}^{\mu }}$, all taken to be 1 for the nonrelativistic Euclidean
oscillator, has non-trivial consequences for the relativistic oscillator.

Kim and Noz \cite{K-N} choose $e^{i\phi _{+}^{\mu }}=~e^{i\phi _{-}^{\mu
}}=1 $ for their relativistic oscillator model, so it follows from (\ref%
{ladder-1}) that $\bar{a}^{0}$ acts as an annihilation operator and $a^{0}$
is the creation operator for the timelike mode, such that the ground state
mode must have $n^{0}\geq 1$. This role reversal between $\bar{a}^{0}$ and $%
a^{0}$ insures that timelike excitations have positive norm, 
\begin{equation}
\langle n^{0}~\mathbf{n}|~n^{0}~\mathbf{n}\rangle =\frac{1}{\left(
n^{0}-1\right) !}\langle 1~\mathbf{0}|\left( \overline{a}^{0}\right)
^{n^{0}-1}\left( a^{0}\right) ^{n^{0}-1}|1~\mathbf{0}\rangle =\langle 1~%
\mathbf{0}|\left( -\eta ^{00}\right) ^{n^{0}-1}|1~\mathbf{0}\rangle =1%
\rule[-0.3cm]{0cm}{0.8cm}%
\end{equation}%
but leads to an indefinite spectrum%
\begin{equation}
\langle n^{0}~\mathbf{n}|K|n^{0}~\mathbf{n}\rangle =\omega \left(
-n^{0}+\tsum_{\mu >0}n^{\mu }+\frac{1}{2}\eta _{\mu \nu }\eta ^{\mu \nu
}\right) .  \label{spec-KN}
\end{equation}%
The requirements on the ground state%
\begin{equation}
\overline{a}^{0}\left\vert 1~\mathbf{0}\right\rangle =0\mbox{\qquad}a^{\mu
}\left\vert 1~\mathbf{0}\right\rangle =0\;,\;\mu >0
\end{equation}%
lead to a set of first order differential equations that reproduce the
ground state solution proposed by Kim and Noz%
\begin{equation}
\psi _{0}\left( x\right) =e^{-t^{2}}e^{-x^{2}/2}=e^{-\left( t^{2}+\mathbf{x}%
^{2}\right) /2}.
\end{equation}

In their study of quark dynamics, Feynman, Kislinger, and Ravndal \cite{FKR}
chose the phases%
\begin{equation}
e^{i\phi _{+}^{0}}=~e^{i\phi _{-}^{0}}=i\mbox{\qquad}e^{i\phi _{+}^{\mu
}}=~e^{i\phi _{-}^{\mu }}=1,\;\mu >0
\end{equation}%
preserving the roles of $\bar{a}^{0}$ as creation operator and $a^{0}$ as
annihilation operator for the timelike mode, under the requirement that $%
n^{0}\leq 0$ so that%
\begin{equation}
\overline{a}^{0}\left\vert n\right\rangle =\sqrt{1-n^{0}}~\left\vert n-%
\mathbf{e}_{0}\right\rangle \mbox{\quad}a^{0}\left\vert n\right\rangle =%
\sqrt{-n^{0}}~\left\vert n+\mathbf{e}_{0}\right\rangle .  \label{a0}
\end{equation}%
Although these states have positive spectrum%
\begin{equation}
\langle -n^{0}~\mathbf{n}|K|-n^{0}~\mathbf{n}\rangle =\omega \left(
n^{0}+\tsum_{\mu >0}n^{\mu }+\frac{1}{2}\eta _{\mu \nu }\eta ^{\mu \nu
}\right)  \label{spec-FKR}
\end{equation}%
they have indefinite norm 
\begin{equation}
\langle n^{0}~\mathbf{n}|~n^{0}~\mathbf{n}\rangle =\frac{1}{\left(
-n^{0}\right) !}\langle 0|\left( a^{0}\right) ^{-n^{0}}\left( \overline{a}%
^{0}\right) ^{-n^{0}}|0\rangle =\langle 0|\left( \eta ^{00}\right)
^{-n^{0}}|0\rangle =\left( -1\right) ^{-n^{0}}  \label{ghost}
\end{equation}%
requiring that the negative norm states (ghosts) be suppressed by exclusion
of excited timelike modes. The first order differential equations $a^{\mu
}\psi _{0}\left( x\right) =0$ lead to the solution proposed in \cite{FKR}, 
\begin{equation}
\psi _{0}\left( x\right) =e^{-x^{2}/2}=e^{-\left( \mathbf{x}%
^{2}-t^{2}\right) /2}  \label{FKR}
\end{equation}%
with some regularization procedure required for normalization.

The ground state energy for uncoupled nonrelativistic oscillators can
usually be found by associating $\frac{1}{2}\hbar \omega $ per degree of
freedom. Although (\ref{spec-KN}) and (\ref{spec-FKR}) indicate a ground
state mass/energy in 4 dimensions of $2\hbar \omega $, the Horwitz-Arshansky
solution exhibits the lower ground state level $\frac{3}{2}\hbar \omega $.
Kastrup \cite{Kastrup} has shown that the choice of Cartesian coordinates
overlooks the singularity at the origin of polar coordinates, implicitly
choosing one solution from a family of harmonic oscillators with different
ground state levels. In the following sections we obtain solutions in polar
coordinates for O(2), O(3), and the spacelike sector of O(2,1), and show
that the ground state level depends on its angular eigenvalue.

\section{Harmonic Oscillators in Polar Coordinates}

In the polar coordinates appropriate to the oscillator problems in $D=2$ and 
$3$ dimensions%
\begin{equation}
\begin{array}{llllc}
x=\rho \cos \phi \mbox{\quad} & y=\rho \sin \phi \mbox{\quad} & \mbox{\quad}
& \mbox{\quad} & \text{O(2)} \\[0.2cm]
x=\rho \cos \phi \sin \theta \mbox{\quad} & y=\rho \sin \phi \sin \theta %
\mbox{\quad} & z=\rho \cos \theta & \mbox{\quad} & \text{O(3)} \\[0.2cm]
x=\rho \cos \phi \cosh \beta \mbox{\quad} & y=\rho \sin \phi \cosh \beta %
\mbox{\quad} & t=\rho \sinh \beta & \mbox{\quad} & \text{O(2,1)}%
\end{array}
\label{pol-coord}
\end{equation}%
the Schrodinger equation takes the form%
\begin{equation}
\left[ -\partial _{\rho }^{2}-\frac{D-1}{\rho }\partial _{\rho }+\frac{1}{%
\rho ^{2}}\mathbf{M}^{2}+\rho ^{2}-\varepsilon \right] \psi =0  \label{schro}
\end{equation}%
where the energy/mass eigenvalue is $E=\frac{\omega }{2}\varepsilon $ and $%
\mathbf{M}^{2}$ is the Casimir operator of the symmetry group, formed from $%
M^{\mu \nu }=x^{\mu }p^{\nu }-x^{\nu }p^{\mu }$, which we notate as%
\begin{equation}
\begin{array}{llllc}
\mbox{\quad} & \mbox{\quad} & M=x^{1}p^{2}-x^{2}p^{1} & \mbox{\quad} & \text{%
O(2)} \\[0.2cm] 
L^{1}=x^{2}p^{3}-x^{3}p^{2}\mbox{\quad} & L^{2}=x^{3}p^{1}-x^{1}p^{3}%
\mbox{\quad} & M=x^{1}p^{2}-x^{2}p^{1} & \mbox{\quad} & \text{O(3)} \\%
[0.2cm] 
A^{1}=x^{0}p^{1}-x^{1}p^{0}\mbox{\quad} & A^{2}=x^{0}p^{2}-x^{2}p^{0}%
\mbox{\quad} & M=x^{1}p^{2}-x^{2}p^{1} & \mbox{\quad} & \text{O(2,1)}%
\end{array}%
\end{equation}%
so that the parameterizations (\ref{pol-coord}) diagonalize the $M^{12}$
angular momentum component%
\begin{equation}
M=x^{1}p^{2}-x^{2}p^{1}=-i\partial _{\phi }.
\end{equation}%
The Casimir operators in these coordinates are%
\begin{equation}
\mathbf{M}^{2}=\left\{ 
\begin{array}{llc}
M^{2}=-\partial _{\phi }^{2} & \mbox{\quad} & \text{O(2)} \\[0.4cm] 
\mathbf{L}^{2}=-\partial _{\theta }^{2}-\dfrac{\cos \theta }{\sin \theta }%
\partial _{\theta }-\dfrac{1}{\sin ^{2}\theta }\partial _{\phi }^{2} & %
\mbox{\quad} & \text{O(3)} \\[0.5cm] 
\mathbf{\Lambda }^{2}=M^{2}-\mathbf{A}^{2}=\partial _{\beta }^{2}+\dfrac{%
\sinh \beta }{\cosh \beta }\partial _{\beta }-\dfrac{1}{\cosh ^{2}\beta }%
\partial _{\phi }^{2} & \mbox{\quad} & \text{O(2,1)}%
\end{array}%
\right.  \label{cas}
\end{equation}%
so assuming a separation of variables%
\begin{equation}
\begin{array}{llc}
\psi \left( \rho ,\phi \right) =R\left( \rho \right) \Phi \left( \phi \right)
& \mbox{\quad} & \text{O(2)} \\[0.4cm] 
\psi \left( \rho ,\theta ,\phi \right) =R\left( \rho \right) F\left( \theta
\right) \Phi \left( \phi \right) & \mbox{\quad} & \text{O(3)} \\[0.5cm] 
\psi \left( \rho ,\beta ,\phi \right) =R\left( \rho \right) G\left( \beta
\right) \Phi \left( \phi \right) & \mbox{\quad} & \text{O(2,1)}%
\end{array}%
\end{equation}%
leads to the common angular function%
\begin{equation}
\Phi \left( \phi \right) =e^{i\Lambda _{1}\phi }  \label{ang-f}
\end{equation}%
allowing the replacement of $M=-i\partial _{\phi }$ in (\ref{cas}) by its
eigenvalue $\Lambda _{1}$. For $D=3$ a second separation of variables,
associated with the eigenvalue equation $\mathbf{M}^{2}\psi =\Lambda
_{2}\psi $ for the Casimir operators, leads to 
\begin{equation}
\left( -\mathbf{M}^{2}+\Lambda _{2}\right) F\left( \theta \right) =\left(
\partial _{\theta }^{2}+\frac{\cos \theta }{\sin \theta }\partial _{\theta }-%
\frac{\Lambda _{1}^{2}}{\sin ^{2}\theta }+\Lambda _{2}\right) F\left( \theta
\right) =0  \label{ang-2}
\end{equation}%
\begin{equation}
\left( \mathbf{M}^{2}-\Lambda _{2}\right) G\left( \beta \right) =\left(
\partial _{\beta }^{2}+\frac{\sinh \beta }{\cosh \beta }\partial _{\beta }+%
\frac{\Lambda _{1}^{2}}{\cosh ^{2}\beta }-\Lambda _{2}\right) G\left( \beta
\right) =0  \label{beta}
\end{equation}%
which may be approached in two inequivalent ways. The first, following the
method applied to the classical central force problems, notes the form of
the first order derivative terms and substitutes%
\begin{equation}
z=\cos \theta \mbox{\qquad}\zeta =\sinh \beta \mbox{\qquad}\Lambda
_{2}=l\left( l+1\right) \mbox{\qquad}\Lambda _{1}=m  \label{sub-4}
\end{equation}%
so that the partial derivative terms for $\theta $ become 
\begin{eqnarray}
\partial _{\theta } &=&-\sin \theta \partial _{z}=-\sqrt{1-z^{2}}\partial
_{z}\mbox{\qquad}\mbox{\qquad}\frac{\cos \theta }{\sin \theta }\partial
_{\theta }=-z\partial _{z} \\[0.2cm]
\partial _{\theta }^{2} &=&\sqrt{1-z^{2}}\partial _{z}\sqrt{1-z^{2}}\partial
_{z}=\left( 1-z^{2}\right) \partial _{z}^{2}-z\partial _{z}
\end{eqnarray}%
and for $\beta $ become%
\begin{eqnarray}
\partial _{\beta } &=&\cosh \beta \partial _{\zeta }=\sqrt{1+\zeta ^{2}}%
\partial _{\zeta }\mbox{\qquad}\mbox{\qquad}\frac{\sinh \beta }{\cosh \beta }%
\partial _{\beta }=\zeta \partial _{\zeta } \\[0.2cm]
\partial _{\beta }^{2} &=&\sqrt{1+\zeta ^{2}}\partial _{\zeta }\sqrt{1+\zeta
^{2}}\partial _{\zeta }=\left( 1+\zeta ^{2}\right) \partial _{\zeta
}^{2}+\zeta \partial _{\zeta }.
\end{eqnarray}%
Writing $F\left( \theta \right) \rightarrow P_{l}^{m}\left( z\right) $ and $%
G\left( \beta \right) \rightarrow \hat{P}_{l}^{m}\left( \zeta \right) $
equations (\ref{ang-2}) and (\ref{beta}) become solutions to the associated
Legendre equation in the respective forms%
\begin{equation}
\left[ \left( 1-z^{2}\right) \partial _{z}^{2}-2z\partial _{z}+l\left(
l+1\right) -\frac{m^{2}}{1-z^{2}}\right] P_{l}^{m}\left( z\right) =0
\label{a-Leg}
\end{equation}%
\begin{equation}
\left[ \left( 1+\zeta ^{2}\right) \partial _{\zeta }^{2}+2\zeta \partial
_{\zeta }-l\left( l+1\right) +\frac{m^{2}}{1+\zeta ^{2}}\right] \hat{P}%
_{l}^{m}\left( \zeta \right) =0.  \label{a-Leg-hat}
\end{equation}%
Notice that (\ref{a-Leg-hat}) can be obtained from (\ref{a-Leg}) by letting%
\begin{equation}
z=i\zeta \rightarrow z^{2}=-\zeta ^{2}\mbox{\qquad}\partial
_{z}^{2}\rightarrow -\partial _{\zeta }^{2}\mbox{\qquad}z\partial
_{z}\rightarrow \zeta \partial _{\zeta }.
\end{equation}%
A second, qualitatively different set of solutions is obtained by
substituting%
\begin{eqnarray}
z &=&\frac{\cos \theta }{\sin \theta }\mbox{\qquad}\mbox{\qquad}%
F_{l}^{m}\left( z\right) =\left( 1+z^{2}\right) ^{\frac{1}{4}}\hat{P}%
_{l}^{m}\left( z\right)  \label{subs_hf} \\[0.2cm]
\zeta &=&\frac{\sinh \beta }{\cosh \beta }\mbox{\qquad}\mbox{\qquad}%
G_{l}^{m}\left( \zeta \right) =\left( 1-\zeta ^{2}\right) ^{\frac{1}{4}%
}P_{l}^{m}\left( \zeta \right)  \label{subs_hf2}
\end{eqnarray}%
so that the partial derivative terms for $\theta $ become 
\begin{eqnarray}
\partial _{\theta } &=&-\frac{1}{\sin ^{2}\theta }\partial _{z}=-\left(
1+z^{2}\right) \partial _{z}\mbox{\qquad}\mbox{\qquad}\frac{\cos \theta }{%
\sin \theta }\partial _{\theta }=-z\left( 1+z^{2}\right) \partial _{z} \\%
[0.2cm]
\partial _{\theta }^{2} &=&\left( 1+z^{2}\right) \partial _{z}\left(
1+z^{2}\right) \partial _{z}=\left( 1+z^{2}\right) ^{2}\partial
_{z}^{2}+\left( 1+z^{2}\right) 2z\partial _{z} \\
\partial _{\theta }^{2}+\cot \theta \partial _{\theta } &=&\left(
1+z^{2}\right) \left[ \left( 1+z^{2}\right) \partial _{z}^{2}+z\partial _{z}%
\right]
\end{eqnarray}%
and for $\beta $ become%
\begin{eqnarray}
\partial _{\beta } &=&\frac{1}{\cosh {}^{2}{}\beta }\partial _{\zeta
}=\left( 1-\zeta ^{2}\right) \partial _{\zeta }\mbox{\qquad}\mbox{\qquad}%
\frac{\sinh \beta }{\cosh \beta }\partial _{\beta }=\zeta \left( 1-\zeta
^{2}\right) \partial _{\zeta } \\[0.2cm]
\partial _{\beta }^{2} &=&\left( 1-\zeta ^{2}\right) \partial _{\zeta
}\left( 1-\zeta ^{2}\right) \partial _{\zeta }=\left( 1-\zeta ^{2}\right) 
\left[ \left( 1-\zeta ^{2}\right) \partial _{\zeta }^{2}-2\zeta \partial
_{\zeta }\right] \\
\partial _{\beta }^{2}+\tanh \beta \partial _{\beta } &=&\left( 1-\zeta
^{2}\right) \left[ \left( 1-\zeta ^{2}\right) \partial _{\zeta }^{2}-\zeta
\partial _{\zeta }\right] .
\end{eqnarray}%
Using%
\begin{eqnarray}
\partial _{z}\left( 1+z^{2}\right) ^{\frac{1}{4}}\hat{P}\left( z\right)
&=&\left( 1+z^{2}\right) ^{\frac{1}{4}}\left[ \partial _{z}+\frac{1}{2}\frac{%
z}{1+z^{2}}\right] \hat{P}\left( z\right) \\[0.2cm]
\partial _{\zeta }\left( 1-\zeta ^{2}\right) ^{\frac{1}{4}}P\left( \zeta
\right) &=&\left( 1-\zeta ^{2}\right) ^{\frac{1}{4}}\left[ \partial _{\zeta
}-\frac{1}{2}\frac{\zeta }{1-\zeta ^{2}}\right] P\left( \zeta \right)
\end{eqnarray}%
we are led to associated Legendre equations 
\begin{eqnarray}
\left[ \left( 1+z^{2}\right) \partial _{z}^{2}+2z\partial _{z}-m\left(
m+1\right) +\frac{l^{2}}{1+z^{2}}\right] \hat{P}_{m}^{l}\left( z\right) &=&0
\label{a-Lag-im} \\
\left[ \left( 1-\zeta ^{2}\right) \partial _{\zeta }^{2}-2\zeta \partial
_{\zeta }+m\left( m+1\right) -\frac{l^{2}}{1-\zeta ^{2}}\right]
P_{m}^{l}\left( \zeta \right) &=&0
\end{eqnarray}%
where the constants $l$ and $m$ have reversed roles with respect to
equations (\ref{a-Leg}) and (\ref{a-Leg-hat}), having been introduced to
satisfy 
\begin{eqnarray}
m\left( m+1\right) &=&\Lambda _{1}^{2}-\frac{1}{4}\mbox{\qquad}%
\longrightarrow \mbox{\qquad}\Lambda _{1}=m+\frac{1}{2}  \label{sub-5} \\
l^{2} &=&\Lambda _{2}+\frac{1}{4}\mbox{\qquad}\longrightarrow \mbox{\qquad}%
\Lambda _{2}=l^{2}-\frac{1}{4}.  \label{sub-6}
\end{eqnarray}%
Comparing (\ref{sub-4}) and (\ref{sub-5}), we write $\Lambda _{1}=m+s$, so
that (\ref{ang-f}) becomes 
\begin{equation}
\Phi \left( \phi \right) =e^{i\Lambda _{1}\phi }=e^{i\left( m+s\right) \phi }
\end{equation}%
where, for $D=3$, the orbital angular momentum is characterized by $s=0,1/2$
and may be integral or half-integral. For $D=2$ we assume that $s$ can be
any real constant.

The remaining radial equations are%
\begin{equation}
\left[ -\partial _{\rho }^{2}-\frac{1}{\rho }\partial _{\rho }+\frac{1}{\rho
^{2}}\Lambda _{1}^{2}+\rho ^{2}-\varepsilon \right] R\left( \rho \right) =0,%
\mbox{\quad}D=2  \label{rad-2}
\end{equation}%
\begin{equation}
\left[ -\partial _{\rho }^{2}-\frac{2}{\rho }\partial _{\rho }+\frac{1}{\rho
^{2}}\Lambda _{2}+\rho ^{2}-\varepsilon \right] R\left( \rho \right) =0,%
\mbox{\quad}D=3.  \label{rad-3}
\end{equation}%
The change of variables 
\begin{equation}
x=\rho ^{2}  \label{sub-1}
\end{equation}%
which entails%
\begin{equation}
\partial _{r}=2x^{\frac{1}{2}}\partial _{x}\mbox{\qquad}\partial
_{r}^{2}=2x^{\frac{1}{2}}\partial _{x}2x^{\frac{1}{2}}\partial
_{x}=4x\partial _{x}^{2}+2\partial _{x},
\end{equation}%
and the substitutions%
\begin{equation}
R\left( \rho \right) =x^{\left( m+s\right) /2}e^{-x/2}L\left( x\right) %
\mbox{\qquad}D=2  \label{sub-2}
\end{equation}%
\begin{equation}
R\left( \rho \right) =x^{\left( m-s\right) /2}e^{-x/2}L\left( x\right) %
\mbox{\qquad}D=3  \label{sub-3}
\end{equation}%
in radial equations (\ref{rad-2}) and (\ref{rad-3}) lead to Laguerre
equations, where in $D=2$, $L\left( x\right) $ satisfies%
\begin{equation}
\left[ x\partial _{x}^{2}+\left( m+s-x+1\right) \partial _{x}+\frac{1}{2}%
\left( \frac{1}{2}\varepsilon -m-s-1\right) \right] L_{n}^{\alpha }\left(
x\right) =0
\end{equation}%
\begin{equation}
\alpha =m+s\mbox{\qquad}n=\frac{1}{2}\left( \frac{1}{2}\varepsilon
-m-s-1\right)
\end{equation}%
and in $D=3$, $L\left( x\right) $ satisfies 
\begin{equation}
\left[ x\partial _{x}^{2}+\left( l-s-x+\frac{3}{2}\right) \partial _{x}^{2}+%
\frac{1}{2}\left( \frac{1}{2}\varepsilon -l+s-\frac{3}{2}\right) \right]
L_{n}^{\alpha }\left( x\right) =0
\end{equation}%
\begin{equation}
\alpha =l-s+\frac{1}{2}\mbox{\qquad}n=\frac{1}{2}\left( \frac{1}{2}%
\varepsilon -l+s-\frac{3}{2}\right) .
\end{equation}%
From $E=\frac{\omega }{2}\varepsilon $ the spectra are given by%
\begin{equation}
E=\omega \left( 2n+m+s+1\right) \mbox{\qquad}D=2  \label{energy}
\end{equation}%
\begin{equation}
E=\omega \left( 2n+l+3/2-s\right) \mbox{\qquad}D=3  \label{energy-3d}
\end{equation}%
and the wavefunctions are%
\begin{eqnarray}
\psi _{nm}^{\text{O(2)},s}\left( \rho ,\phi \right) &=&A_{nm}e^{-\rho
^{2}/2}\rho ^{m+s}L_{n}^{m+s}\left( \rho ^{2}\right) e^{i\left( m+s\right)
\phi }  \label{solution_1} \\[0.08in]
\psi _{nlm}^{\text{O(3)},s=0}\left( \rho ,\theta ,\phi \right)
&=&A_{nlm}e^{-\rho ^{2}/2}\rho ^{l}L_{n}^{l+\frac{1}{2}}\left( \rho
^{2}\right) P_{l}^{m}\left( \cos \theta \right) e^{im\phi }  \label{solu-2}
\\[0.08in]
\psi _{nlm}^{\text{O(3)},s=\frac{1}{2}}\left( \rho ,\theta ,\phi \right)
&=&A_{nlm}e^{-\rho ^{2}/2}\rho ^{l-\frac{1}{2}}L_{n}^{l}\left( \rho
^{2}\right) \frac{\hat{P}_{m}^{l}\left( \cot \theta \right) }{\sqrt{%
\left\vert \sin \theta \right\vert }}e^{i\left( m+\frac{1}{2}\right) }
\label{wf-half} \\[0.08in]
\psi _{nlm}^{\text{O(2,1)},s=0}\left( \rho ,\beta ,\phi \right)
&=&A_{nlm}e^{-\rho ^{2}/2}\rho ^{l}L_{n}^{l+\frac{1}{2}}\left( \rho
^{2}\right) \hat{P}_{l}^{m}\left( \sinh \beta \right) e^{im\phi }
\label{solu-4} \\[0.08in]
\psi _{nlm}^{\text{O(2,1)},s=\frac{1}{2}}\left( \rho ,\beta ,\phi \right)
&=&A_{nlm}e^{-\rho ^{2}/2}\rho ^{l-\frac{1}{2}}L_{n}^{l}\left( \rho
^{2}\right) \frac{P_{m}^{l}\left( \tanh \beta \right) }{\sqrt{\cosh \beta }}%
e^{i\left( m+\frac{1}{2}\right) \phi }.  \label{solu-5}
\end{eqnarray}%
Using the properties $L_{\beta }^{\alpha }=0$ for $\beta <0$ and $%
L_{0}^{\alpha }=P_{0}^{0}=1$, the wavefunctions with eigenvalues $n=l=m=0$ are
summarized as%
\begin{equation}
\psi _{0}^{\text{O(2)},s}\left( \rho ,\phi \right) =A_{0}e^{-\rho
^{2}/2}\left( \rho e^{i\phi }\right) ^{s}=A_{0}e^{-\left( x^{2}+y^{2}\right)
/2}\left( x^{2}+y^{2}\right) ^{s/2}e^{is\arctan \left( \frac{y}{x}\right) }
\label{G2}
\end{equation}%
\begin{equation}
\psi _{0}^{\text{O(3)},s}\left( \rho ,\theta ,\phi \right) =A_{0}e^{-\rho
^{2}/2}\frac{e^{is\phi }}{\left( \rho \left\vert \sin \theta \right\vert
\right) ^{s}}=A_{0}e^{-\left( x^{2}+y^{2}+z^{2}\right) /2}\frac{e^{is\arctan
\left( \frac{y}{x}\right) }}{\left( x^{2}+y^{2}\right) ^{s/2}}
\end{equation}%
\begin{equation}
\psi _{0}^{\text{O(2,1)},s}\left( \rho ,\beta ,\phi \right) =A_{0}e^{-\rho
^{2}/2}\frac{e^{is\phi }}{\left( \rho \cosh \beta \right) ^{s}}%
=A_{0}e^{-\left( x^{2}+y^{2}-t^{2}\right) /2}\frac{e^{is\arctan \left( \frac{%
y}{x}\right) }}{\left( x^{2}+y^{2}\right) ^{s/2}}  \label{G4}
\end{equation}%
and so, as expected, are separable in Cartesian coordinates only for $s=0$,
in which case they recover the standard solutions expressed as products of
one dimensional oscillators. In particular, the $s=0$ ground state for
O(2,1) is precisely the state proposed by Feynman, Kislinger, and Ravndal.

\section{Number Representation in Polar Coordinates}

A number representation appropriate to the solutions (\ref{solution_1}) to (%
\ref{solu-5}) consists of polar creation/annihilation operators that act on
polar eigenstates of the total mode number $N$ and the symmetry operators $%
\mathbf{M}^{2}$ and $M$ to produce new polar eigenstates. The resulting
representation will be equivalent to the standard Cartesian Fock space if
the polar eigenstates are unitarily connected to the Cartesian number
states, in which case they can be found by expressing $N,\mathbf{M}^{2}$ and 
$M$ in terms of $\bar{a}^{\mu }$ and $a^{\mu }$ and diagonalizing the
resulting operators. We consider the Cartesian multiplet $\varphi _{1}$ of
first excited states as arising from the action of the vector multiplet of
creation operators on the ground state $\varphi _{0}$. Thus, in $D=2$ the
vector operator multiplet takes $\varphi _{0}$ to $\varphi _{1}$ as 
\begin{equation}
\varphi _{1}=\left( 
\begin{array}{c}
\varphi _{10} \\ 
\varphi _{01}%
\end{array}%
\right) =\left( 
\begin{array}{c}
\bar{a}^{1}\varphi _{0} \\ 
\bar{a}^{2}\varphi _{0}%
\end{array}%
\right) =\left( 
\begin{array}{c}
\bar{a}^{1} \\ 
\bar{a}^{2}%
\end{array}%
\right) \varphi _{0}.
\end{equation}%
Using (\ref{cr/an}) to replace $x^{\mu }$ and $p^{\mu }$ with $\bar{a}^{\mu
} $ and $a^{\mu }$, the angular momentum operator%
\begin{equation}
M=x^{1}p^{2}-x^{2}p^{1}=-i\left( \bar{a}^{1}a^{2}-\bar{a}^{2}a^{1}\right)
\end{equation}%
is seen to act on $\varphi _{1}$ as%
\begin{equation}
M\varphi _{1}=-i\left( \bar{a}^{1}a^{2}-\bar{a}^{2}a^{1}\right) \left( 
\begin{array}{c}
\bar{a}^{1}\varphi _{0} \\ 
\bar{a}^{2}\varphi _{0}%
\end{array}%
\right) =\left( 
\begin{array}{c}
i\bar{a}^{2}\varphi _{0} \\ 
-i\bar{a}^{1}\varphi _{0}%
\end{array}%
\right) =\left( 
\begin{array}{cc}
0 & i \\ 
-i & 0%
\end{array}%
\right) \varphi _{1}
\end{equation}%
and so has eigenvalues $\pm 1$ on eigenstates%
\begin{equation}
\tilde{\varphi}_{1}=\frac{1}{\sqrt{2}}\left( 
\begin{array}{c}
\varphi _{10}+i\varphi _{01} \\ 
-\varphi _{10}+i\varphi _{01}%
\end{array}%
\right) =\frac{1}{\sqrt{2}}\left( 
\begin{array}{c}
\bar{a}_{+} \\ 
-\bar{a}_{-}%
\end{array}%
\right) \varphi _{0}
\end{equation}%
where the polar creation/annihilation operators%
\begin{equation}
a_{\pm }=\frac{1}{\sqrt{2}}\left( a^{1}\pm ia^{2}\right) \mbox{\qquad}\bar{a}%
_{\pm }=\frac{1}{\sqrt{2}}\left( \bar{a}^{1}\pm i\bar{a}^{2}\right)
\label{a+-}
\end{equation}%
commute among themselves except for%
\begin{equation}
\left[ a_{+},\bar{a}_{-}\right] =\left[ a_{-},\bar{a}_{+}\right] =1.
\label{aa*}
\end{equation}%
Since $a^{1}$ and $a^{2}$ commute with $\bar{a}^{0}$ and $\bar{a}^{3}$, the
operators defined in (\ref{a+-}) similarly diagonalize $M$ in $D=3$, with
eigenvalue 0 on the states $\bar{a}^{0}\varphi _{0}$ and $\bar{a}^{3}\varphi
_{0}$.

\subsection{Number representation for $D=2$}

Because $\mathbf{M}^{2}=\left( M\right) ^{2}$ in $D=2$, operators (\ref{a+-}%
) are sufficient to fully characterize the O(2) oscillator. From the four
available products 
\begin{eqnarray}
\bar{a}_{+}a_{+} &=&\frac{1}{2}\left( \bar{a}^{1}+i\bar{a}^{2}\right) \left(
a^{1}+ia^{2}\right) =\frac{1}{2}\left( N^{1}-N^{2}+i\bar{a}^{2}a^{1}+i\bar{a}%
^{1}a^{2}\right) \\
\bar{a}_{-}a_{-} &=&\frac{1}{2}\left( \bar{a}^{1}-i\bar{a}^{2}\right) \left(
a^{1}-ia^{2}\right) =\frac{1}{2}\left( N^{1}-N^{2}-i\bar{a}^{2}a^{1}-i\bar{a}%
^{1}a^{2}\right) \\
\bar{a}_{+}a_{-} &=&\frac{1}{2}\left( \bar{a}^{1}+i\bar{a}^{2}\right) \left(
a^{1}-ia^{2}\right) =\frac{1}{2}\left( N^{1}+N^{2}+i\bar{a}^{2}a^{1}-i\bar{a}%
^{1}a^{2}\right) \\
\bar{a}_{-}a_{+} &=&\frac{1}{2}\left( \bar{a}^{1}-i\bar{a}^{2}\right) \left(
a^{1}+ia^{2}\right) =\frac{1}{2}\left( N^{1}+N^{2}-i\bar{a}^{2}a^{1}+i\bar{a}%
^{1}a^{2}\right)
\end{eqnarray}%
we may form the symmetric Hermitian combinations 
\begin{equation}
N=\bar{a}_{+}a_{-}+\bar{a}_{-}a_{+}=N^{1}+N^{2}  \label{N}
\end{equation}%
\begin{equation}
\Delta =\bar{a}_{+}a_{+}+\bar{a}_{-}a_{-}=N^{1}-N^{2}=\bar{a}^{1}a^{1}-\bar{a%
}^{2}a^{2}  \label{Q+}
\end{equation}%
and the antisymmetric Hermitian combinations%
\begin{equation}
M=\bar{a}_{+}a_{-}-\bar{a}_{-}a_{+}=-i\left( \bar{a}^{1}a^{2}-\bar{a}%
^{2}a^{1}\right)  \label{M}
\end{equation}%
\begin{equation}
Q=-i\left( \bar{a}_{+}a_{+}-\bar{a}_{-}a_{-}\right) =\bar{a}^{1}a^{2}+\bar{a}%
^{2}a^{1}.  \label{Q-}
\end{equation}%
In Cartesian coordinates, the maximal set of commuting operators is $\left\{
N^{1},N^{2}\right\} $, and from these we construct the Hamiltonian. Using%
\begin{eqnarray}
\left[ M,N^{1}\right] &=&-i\left[ \bar{a}^{1}a^{2}-a^{1}\bar{a}^{2},\bar{a}%
^{1}a^{1}\right] =i\left( \bar{a}^{1}a^{2}+\bar{a}^{2}a^{1}\right) =iQ \\
\left[ M,N^{2}\right] &=&-i\left[ \bar{a}^{1}a^{2}-a^{1}\bar{a}^{2},\bar{a}%
^{2}a^{2}\right] =-i\left( \bar{a}^{1}a^{2}+\bar{a}^{2}a^{1}\right) -iQ
\end{eqnarray}%
we confirm that angular momentum commutes with the total mode number%
\begin{equation}
\left[ M,N\right] =\left[ M,N^{1}\right] +\left[ M,N^{2}\right] =0
\label{M-H}
\end{equation}%
but since $N^{1}$ and $N^{2}$ do not commute with $M$, they are not
separately observable in the polar representation.

Since $N$ is a positive operator, we must address the problem of negative
energy states. From (\ref{energy}) the energy of $n=0$ states can become
negative if $m+s<-1$. For eigenvalues $n\geq 0$ and $m+s\geq 0$, the
wavefunctions (\ref{solution_1}) are made orthonormal by taking the
normalization to be%
\begin{equation}
A_{nm}=\frac{\left( -1\right) ^{n}}{\sqrt{\int_{0}^{2\pi }d\phi
\int_{0}^{\infty }\rho d\rho ~\left\vert \psi _{nm}^{\text{O(2)},s}\left(
\rho ,\phi \right) \right\vert ^{2}}}=\left( -1\right) ^{n}\sqrt{\frac{%
\Gamma \left( n+1\right) }{\pi \Gamma \left( n+m+s+1\right) }}
\label{norm-2}
\end{equation}%
but for states with $n=0$ and $m+s<0$, this becomes%
\begin{equation}
A_{0m}=\frac{1}{\sqrt{\int_{0}^{2\pi }d\phi \int_{0}^{\infty }\rho d\rho
~\left\vert \psi _{0m}^{\text{O(2)},s}\left( \rho ,\phi \right) \right\vert
^{2}}}=\frac{1}{\sqrt{\pi \int_{0}^{\infty }dx~e^{-x/2}x^{m+s}}}\underset{%
m+s<0}{\longrightarrow }0
\end{equation}%
eliminating negative energy states $\psi _{0m}$ for $m+s<0$. The general
normalized wavefunction is then%
\begin{equation}
\psi _{nm}\left( r,\phi \right) =\left\{ 
\begin{array}{ll}
\left( -1\right) ^{n}\sqrt{\frac{\Gamma \left( n+1\right) }{\pi \Gamma
\left( n+m+s+1\right) }}r^{m+s}e^{-\frac{r^{2}}{2}}L_{n}^{m+s}\left(
r^{2}\right) e^{i\left( m+s\right) \phi } & ,~n\geq 0,n+m+s\geq 0 \\ 
0 & ,~\text{ otherwise }%
\end{array}%
\right.  \label{wavefunction}
\end{equation}%
with positive definite total energy. We may satisfy the requirement%
\begin{equation}
n+m+s\geq 0  \label{ms_pos}
\end{equation}%
by taking $0\leq s<1$ and $m\geq 0$. From (\ref{hamil}) the Hamiltonian in $%
D=2$ is $H=\omega \left( N+1\right) $ so comparing with (\ref{energy}) we
find%
\begin{equation}
N=2n+m+s  \label{total-modes}
\end{equation}%
where to avoid confusion with the principal quantum number $n$, we use $N$
to represent both the total mode number operator and its eigenvalue. Using
the commutation relations 
\begin{equation}
\left[ N,a_{\pm }\right] =-a_{\pm }\mbox{\qquad}\left[ N,\bar{a}_{\pm }%
\right] =\bar{a}_{\pm }  \label{Na}
\end{equation}%
\begin{equation}
\left[ M,a_{\pm }\right] =\pm a_{\pm }\mbox{\qquad}\left[ M,\bar{a}_{\pm }%
\right] =\pm \bar{a}_{\pm }.  \label{Ma}
\end{equation}%
we compare%
\begin{eqnarray}
Na_{+}\psi _{nm} &=&a_{+}\left( N-1\right) \psi _{nm}=\left( 2n+m+s-1\right)
a_{+}\psi _{nm} \\
Ma_{+}\psi _{nm} &=&\left( m+s+1\right) a_{+}\psi _{nm}
\end{eqnarray}%
with%
\begin{eqnarray}
N\psi _{n-1,m+1} &=&\left( 2n+m+s-1\right) \psi _{n-1,m+1} \\
M\psi _{n-1,m+1} &=&\left( m+s+1\right) \psi _{n-1,m+1}
\end{eqnarray}%
and conclude that%
\begin{equation}
a_{+}\psi _{nm}=C_{nm}^{+}\psi _{n-1,m+1}
\end{equation}%
where $C_{nm}^{+}$ is a complex coefficient, with norm found from 
\begin{equation}
\left\Vert a_{+}\psi _{nm}\right\Vert ^{2}=\left\langle \psi
_{nm}\right\vert \bar{a}_{-}a_{+}\left\vert \psi _{nm}\right\rangle
=\left\vert C_{nm}^{+}\right\vert ^{2}\left\langle \psi _{n-1,m+1}|\psi
_{n-1,m+1}\right\rangle =\left\vert C_{nm}^{+}\right\vert ^{2}~~.
\end{equation}%
Similarly comparing,%
\begin{eqnarray}
Na_{-}\psi _{nm} &=&\left( 2n+m+s-1\right) a_{-}\psi _{nm} \\
Ma_{-}\psi _{nm} &=&\left( m+s-1\right) a_{-}\psi _{nm}
\end{eqnarray}%
\begin{eqnarray}
N\bar{a}_{+}\psi _{nm} &=&\left( 2n+m+s+1\right) a_{+}\psi _{nm} \\
M\bar{a}_{+}\psi _{nm} &=&\left( m+s+1\right) \bar{a}_{+}\psi _{nm}
\end{eqnarray}%
\begin{eqnarray}
N\bar{a}_{-}\psi _{nm} &=&\left( 2n+m+s+1\right) \bar{a}_{-}\psi _{nm} \\
M\bar{a}_{-}\psi _{nm} &=&\left( m+s-1\right) \bar{a}_{-}\psi _{nm}
\end{eqnarray}%
with%
\begin{eqnarray}
N\psi _{n,m-1} &=&\left( 2n+\left( m-1\right) \right) \psi _{n,m-1}=\left(
2n+s+m-1\right) \psi _{n,m-1} \\
M\psi _{n,m-1} &=&\left( m-1\right) \psi _{n,m-1}
\end{eqnarray}%
\begin{eqnarray}
N\psi _{n,m+1} &=&\left[ 2n+s+\left( m+1\right) \right] \psi _{n,m+1}=\left(
2n+s+m+1\right) \psi _{n,m+1} \\
M\psi _{n,m+1} &=&\left( m+s+1\right) \psi _{n,m+1}
\end{eqnarray}%
\begin{eqnarray}
N\psi _{n+1,m-1} &=&\left[ 2\left( n+1\right) +s+\left( m-1\right) \right]
\psi _{n+1,m-1}=\left( 2n+s+m+1\right) \psi _{n+1,m-1} \\
M\psi _{n+1,m-1} &=&\left( m+s-1\right) \psi _{n+1,m-1}
\end{eqnarray}%
leads to%
\begin{equation}
a_{-}\psi _{nm}=C_{nm}^{-}\psi _{nm-1}\mbox{\qquad}\bar{a}_{+}\psi _{nm}=%
\bar{C}_{nm}^{+}\psi _{nm+1}\mbox{\qquad}\bar{a}_{-}\psi _{nm}=\bar{C}%
_{nm}^{-}\psi _{n+1m-1}.
\end{equation}%
We eliminate two coefficients 
\begin{equation}
\left\vert \bar{C}_{nm}^{-}\right\vert ^{2}=\left\vert C_{nm}^{+}\right\vert
^{2}+1\mbox{\qquad}\left\vert \bar{C}_{nm}^{+}\right\vert ^{2}=\left\vert
C_{nm}^{-}\right\vert ^{2}+1  \label{aa*-1}
\end{equation}%
using the commutation relations (\ref{aa*}). Solving%
\begin{equation}
2n+m+s=\left\langle nm\right\vert N\left\vert nm\right\rangle =\left\langle
nm\right\vert \bar{a}_{+}a_{-}+\bar{a}_{-}a_{+}\left\vert nm\right\rangle
=\left\vert C_{nm}^{-}\right\vert ^{2}+\left\vert C_{nm}^{+}\right\vert
^{2}\geq 0,  \label{pos-spec}
\end{equation}%
which requires that the total mode number be positive, together with%
\begin{equation}
m+s=\left\langle nm\right\vert M\left\vert nm\right\rangle =\left\langle
nm\right\vert \bar{a}_{+}a_{-}-\bar{a}_{-}a_{+}\left\vert nm\right\rangle
=\left\vert C_{nm}^{-}\right\vert ^{2}-\left\vert C_{nm}^{+}\right\vert ^{2}
\end{equation}%
and taking the coefficients to be real%
\begin{eqnarray}
C_{nm}^{+} &=&\sqrt{n}\mbox{\qquad}\mbox{\qquad}\bar{C}_{nm}^{-}=\sqrt{n+1}
\\
C_{nm}^{-} &=&\sqrt{n+m+s}\mbox{\qquad}\bar{C}_{nm}^{+}=\sqrt{n+m+s+1}
\end{eqnarray}%
we write the actions of the ladder operators as%
\begin{eqnarray}
a_{+}\psi _{nm} &=&\sqrt{n}~\psi _{n-1,m+1}\mbox{\qquad}\bar{a}_{-}\psi
_{nm}=\sqrt{n+1}~\psi _{n+1,m-1}  \label{actions-1} \\
a_{-}\psi _{nm} &=&\sqrt{n+m+s}~\psi _{n,m-1}\mbox{\qquad}\bar{a}_{+}\psi
_{nm}=\sqrt{n+m+s+1}~\psi _{n,m+1}~.  \label{actions-2}
\end{eqnarray}%
Special care must be taken with the ground state (\ref{G2}) because (\ref%
{actions-1}) and (\ref{actions-2}) lead to%
\begin{equation}
a_{+}\psi _{0}=0\mbox{\qquad}a_{-}\psi _{0}=\sqrt{s}~\psi _{0,-1}
\label{act_on_ground-1}
\end{equation}%
or equivalently 
\begin{equation}
a^{1}\psi _{0}=\sqrt{\frac{s}{2}}\psi _{0,-1}\mbox{\qquad}a^{2}\psi _{0}=i%
\sqrt{\frac{s}{2}}\psi _{0,-1}  \label{act_on_ground-2}
\end{equation}%
suggesting a negative energy state. However, the well-defined, non-zero
function $\psi _{0,-1}$ is non-normalizable and by (\ref{wavefunction}) does
not correspond to any state in the Fock space. We interpret the action of $%
a_{-}$ in (\ref{act_on_ground-1}) as taking the ground state to a
non-observable function which must be taken account in calculations such as 
\begin{equation}
N\psi _{0}=\left( \bar{a}_{+}a_{-}+\bar{a}_{-}a_{+}\right) \psi _{0}=\sqrt{s}%
\bar{a}_{+}\psi _{0,-1}=\sqrt{s}\sqrt{-1+s+1}\psi _{0}=s\psi _{0}
\label{N_grnd}
\end{equation}%
but is effectively annihilated at the end of calculations.

We may construct excited states from the ground state as%
\begin{equation}
\zeta _{\alpha \beta }=\frac{1}{N_{\alpha \beta }}\left( \bar{a}_{+}\right)
^{\alpha }\left( \bar{a}_{-}\right) ^{\beta }\psi _{0}  \label{gen-states}
\end{equation}%
with normalization coefficient $N_{\alpha \beta }$. It follows from (\ref%
{actions-1}) that%
\begin{equation}
\left( \bar{a}_{-}\right) ^{\beta }\psi _{0}=\sqrt{\beta !}~\psi _{\beta
,-\beta }
\end{equation}%
and from (\ref{actions-2}) that%
\begin{equation}
\left( \bar{a}_{+}\right) ^{\alpha }\psi _{\beta ,-\beta }=\sqrt{\frac{%
\Gamma \left( s+\alpha +1\right) }{\Gamma \left( s+1\right) }}~\psi _{\beta
,-\beta +\alpha }
\end{equation}%
and so we take%
\begin{equation}
N_{\alpha \beta }=\sqrt{\beta !\frac{\Gamma \left( s+\alpha +1\right) }{%
\Gamma \left( s+1\right) }}  \label{Nab}
\end{equation}%
which reduces to $\sqrt{\alpha !\beta !}$ in the case $s=0$. Operating on
these states with the total mode operator (\ref{N})%
\begin{equation}
N\zeta _{\alpha \beta }=\frac{1}{N_{\alpha \beta }}\left( \bar{a}_{+}a_{-}+%
\bar{a}_{-}a_{+}\right) \left( \bar{a}_{+}\right) ^{\alpha }\left( \bar{a}%
_{-}\right) ^{\beta }\psi _{0}
\end{equation}%
with the commutation relations (\ref{aa*}) and the identity%
\begin{equation}
\left[ B,A\right] =c\longrightarrow \left[ B,A^{n}\right] =cnA^{n-1}
\end{equation}%
we calculate%
\begin{eqnarray}
\bar{a}_{+}a_{-}\left( \bar{a}_{+}\right) ^{\alpha }\left( \bar{a}%
_{-}\right) ^{\beta }\psi _{0} &=&\bar{a}_{+}\left[ \left( \bar{a}%
_{+}\right) ^{\alpha }a_{-}+\alpha \left( \bar{a}_{+}\right) ^{\alpha -1}%
\right] \left( \bar{a}_{-}\right) ^{\beta }\psi _{0} \\
&=&\left[ \left( \bar{a}_{+}\right) ^{\alpha +1}\left( \bar{a}_{-}\right)
^{\beta }a_{-}+\alpha \left( \bar{a}_{+}\right) ^{\alpha }\left( \bar{a}%
_{-}\right) ^{\beta }\right] \psi _{0} \\
&=&\left( \bar{a}_{+}\right) ^{\alpha }\left( \bar{a}_{-}\right) ^{\beta
}\left( \bar{a}_{+}a_{-}+\alpha \right) \psi _{0}
\end{eqnarray}%
\begin{eqnarray}
\bar{a}_{-}a_{+}\left( \bar{a}_{+}\right) ^{\alpha }\left( \bar{a}%
_{-}\right) ^{\beta }\psi _{0} &=&\left( \bar{a}_{+}\right) ^{\alpha }\bar{a}%
_{-}a_{+}\left( \bar{a}_{-}\right) ^{\beta }\psi _{0} \\
&=&\left( \bar{a}_{+}\right) ^{\alpha }\bar{a}_{-}\left[ \left( \bar{a}%
_{-}\right) ^{\beta }a_{+}+\beta \left( \bar{a}_{-}\right) ^{\beta -1}\right]
\psi _{0} \\
&=&\left( \bar{a}_{+}\right) ^{\alpha }\left( \bar{a}_{-}\right) ^{\beta
}\left( \bar{a}_{-}a_{+}+\beta \right) \psi _{0}
\end{eqnarray}%
which combine to%
\begin{equation}
N\zeta _{\alpha \beta }=\frac{1}{N_{\alpha \beta }}\left( \bar{a}_{+}\right)
^{\alpha }\left( \bar{a}_{-}\right) ^{\beta }\left( \alpha +\beta +\bar{a}%
_{+}a_{-}+\bar{a}_{-}a_{+}\right) \psi _{0}
\end{equation}%
and using (\ref{N_grnd}) we show that 
\begin{equation}
N\zeta _{\alpha \beta }=\left( \alpha +\beta +s\right) \zeta _{\alpha \beta }
\label{tot-mod}
\end{equation}%
so that the states $\zeta _{\alpha \beta }$ have total mode number given by%
\begin{equation}
N=\alpha +\beta +s=\alpha +\beta +N_{\text{ground~state}}~.  \label{N_again}
\end{equation}%
A similar calculation using (\ref{M}) leads to 
\begin{equation}
M\zeta _{\alpha \beta }=\left( \alpha -\beta +s\right) \zeta _{\alpha \beta }
\end{equation}%
so that the states $\zeta _{\alpha \beta }$ have angular momentum%
\begin{equation}
M=\alpha -\beta +s=\alpha -\beta +M_{\text{ground~state}}\mbox{\qquad}%
m=\alpha -\beta ~.  \label{tot-am}
\end{equation}%
Comparing (\ref{tot-mod}) with (\ref{total-modes}) we see that%
\begin{equation}
\alpha +\beta =2n+m
\end{equation}%
which combines with (\ref{tot-am}) and (\ref{N_again}) to provide
expressions for the principal quantum number $n$ and the integer part of the
angular momentum $m$%
\begin{equation}
n=\frac{1}{2}\left( \alpha +\beta -m\right) =\beta \mbox{\qquad}m=\alpha
-\beta =N-s-2\beta  \label{params-1}
\end{equation}%
and fixes $\alpha $ as%
\begin{equation}
\alpha =2n+m-\beta =n+m=N-s-m.  \label{params-2}
\end{equation}%
Thus, the states $\zeta _{\alpha \beta }$ defined in (\ref{gen-states}) can
be identified with explicit solutions $\psi _{n,m}^{\left( N\right) }$
through%
\begin{equation}
\zeta _{\alpha \beta }=\psi _{\beta ,\alpha -\beta }^{\left( \alpha +\beta
+s\right) }\mbox{\qquad}\mbox{\qquad}\psi _{n,m}^{\left( N\right) }=\zeta
_{N-s-m,n}  \label{xi-psi}
\end{equation}%
for which $n=\beta =0,1,...,N-s$ characterizes
the $\left( N-s+1\right) $-fold multiplicity of states with mode number $N$.
Equivalently, the multiplicity can be enumerated by the angular momentum $m$%
, and for $s=0$, this simple multiplicity structure is identical to the
Cartesian picture in which there are $N+1$ ways to build a state of total
mode number $N$ from a pair of one dimensional oscillators.

Bars has recently observed \cite{Bars} that the harmonic oscillator
Hamiltonian in $D$ dimensions possesses a symmetry generated by the products 
$\bar{a}^{\mu }a^{\nu }$ of the ladder operators. Because such products
replace one $\nu $-mode of the oscillator with one $\mu $-mode, the total
mode number, and therefore the total mass/energy, is conserved. The
traceless part of the generators 
\begin{equation}
J^{\mu \nu }=\bar{a}^{\mu }a^{\nu }-\frac{1}{D}\eta ^{\mu \nu }\eta
_{\lambda \rho }\bar{a}^{\lambda }a^{\rho }
\end{equation}%
generates an SU($D-1,1$) or SU($D$) dynamical symmetry of the Hamiltonian,
while the trace%
\begin{equation}
\eta _{\lambda \rho }\bar{a}^{\lambda }a^{\rho }=\tsum_{\mu }\eta _{\mu \mu
}N^{\mu }=N
\end{equation}%
is the total mode number and differs from the Hamiltonian by a c-number. The
antisymmetric part of the generators 
\begin{equation}
\frac{1}{2}\left( J^{\mu \nu }-J^{\nu \mu }\right) =M^{\mu \nu }=\bar{a}%
^{\mu }a^{\nu }-\bar{a}^{\nu }a^{\mu }
\end{equation}%
generates the SO($D-1,1$) or SO($D$) symmetry of the Hamiltonian. Bars
argues that harmonic oscillator states should belong to representations of
the SU dynamical symmetry as well as to representations of the SO symmetry,
imposing additional constraints on admissible solutions.

For the Cartesian ladder operators in two dimensions, the traceless operator
is%
\begin{equation}
J=\left[ 
\begin{array}{cc}
\frac{1}{2}\left( \bar{a}^{1}a^{1}-\bar{a}^{2}a^{2}\right) & \bar{a}^{1}a^{2}
\\ 
\bar{a}^{2}a^{1} & -\frac{1}{2}\left( \bar{a}^{1}a^{1}-\bar{a}%
^{2}a^{2}\right)%
\end{array}%
\right]
\end{equation}%
with antisymmetric part equal to the angular momentum operator%
\begin{equation}
M^{12}=\bar{a}^{1}a^{2}-\bar{a}^{2}a^{1}=iM
\end{equation}%
and symmetric part given by%
\begin{equation}
S=\frac{1}{2}\left( J+J^{\intercal }\right) =\tfrac{1}{2}\left[ 
\begin{array}{cc}
\bar{a}^{1}a^{1}-\bar{a}^{2}a^{2} & \bar{a}^{1}a^{2}+\bar{a}^{2}a^{1} \\ 
\bar{a}^{1}a^{2}+\bar{a}^{2}a^{1} & -\bar{a}^{1}a^{1}+\bar{a}^{2}a^{2}%
\end{array}%
\right] =\tfrac{1}{2}\left[ 
\begin{array}{cc}
\Delta & iQ \\ 
iQ & -\Delta%
\end{array}%
\right]  \label{symm_part}
\end{equation}%
where we use (\ref{Q+}) and (\ref{Q-}). Directly calculating%
\begin{equation}
\left[ \tfrac{1}{2}M,\tfrac{1}{2}\Delta \right] =\tfrac{1}{4}\left[ M,N^{1}%
\right] -\tfrac{1}{4}\left[ M,N^{2}\right] =i\tfrac{1}{2}Q  \label{c1}
\end{equation}%
\begin{equation}
\left[ \tfrac{1}{2}\Delta ,\tfrac{1}{2}Q\right] =-i\tfrac{1}{4}\left[ \bar{a}%
_{+}a_{+}+\bar{a}_{-}a_{-},\bar{a}_{+}a_{+}-\bar{a}_{-}a_{-}\right] =\tfrac{1%
}{2}i\left[ \bar{a}_{+}a_{+},\bar{a}_{-}a_{-}\right] =i\tfrac{1}{2}M
\label{c3}
\end{equation}%
\begin{equation}
\left[ \tfrac{1}{2}Q,\tfrac{1}{2}M\right] =-i\tfrac{1}{4}\left[ \bar{a}%
^{1}a^{2}+\bar{a}^{2}a^{1},\bar{a}^{1}a^{2}-\bar{a}^{2}a^{1}\right] =\tfrac{1%
}{2}i\left[ \bar{a}^{1}a^{2},\bar{a}^{2}a^{1}\right] =i\tfrac{1}{2}\Delta
\label{c2}
\end{equation}%
we verify that the three independent operators $\left\{ \frac{1}{2}M,\frac{1%
}{2}\Delta ,\frac{1}{2}Q\right\} $ satisfy the SU(2) algebra. Equation (\ref%
{M-H}) confirms that $M$ commutes with total mode number $N$ --- similarly,%
\begin{equation}
\left[ N,\Delta \right] =\left[ N^{1}+N^{2},N^{1}-N^{2}\right] =0
\end{equation}%
\begin{equation}
\left[ N,Q\right] =\left[ \bar{a}^{1}a^{1}+\bar{a}^{2}a^{2},\bar{a}^{1}a^{2}+%
\bar{a}^{2}a^{1}\right] =0
\end{equation}%
so this SU(2) is indeed a symmetry of the Hamiltonian. In Cartesian
coordinates, the operator $\Delta $ is chosen to be observable, while in
polar coordinates the operator $M$ is observable. Comparison of (\ref%
{act_on_ground-2}) and (\ref{symm_part}) however indicates that the SU(2)
symmetry is spontaneously broken for the states (\ref{wavefunction}) except
in the case that $s=0$. Moreover, expanding the SU(2) Casimir operator in
terms of the creation and annihilation operators, (\ref{Q+}) --- (\ref{Q-})
lead to%
\begin{equation}
\left( \tfrac{1}{2}M\right) ^{2}+\left( \tfrac{1}{2}\Delta \right)
^{2}+\left( \tfrac{1}{2}Q\right) ^{2}=\tfrac{1}{4}N\left( N+2\right) =\tfrac{%
1}{2}N\left( \tfrac{1}{2}N+1\right) .
\end{equation}%
Since the Casimir eigenvalue $\tfrac{1}{2}N$ of a unitary representation of
SU(2) must be integral or half-integral, the $s\neq 0$ solutions appear to
violate unitarity \cite{Bars-comment}. A more detailed study of the
unitarity of the explicit solutions will be presented in a subsequent paper.

To verify that the solutions (\ref{wavefunction}) form the basis for a
representation of the operator algebra, we express the creation/annihilation
operators in polar coordinates. Combining (\ref{cr/an}) and (\ref{a+-}) as%
\begin{eqnarray}
a_{\pm } &=&\frac{1}{2}\left[ \left( x+\partial _{x}\right) \pm i\left(
y+\partial _{y}\right) \right] =\frac{1}{2}\left[ x\pm iy+\left( \partial
_{x}\pm i\partial _{y}\right) \right] \\
\bar{a}_{\pm } &=&\frac{1}{2}\left[ \left( x-\partial _{x}\right) \pm
i\left( y-\partial _{y}\right) \right] =\frac{1}{2}\left[ x\pm iy-\left(
\partial _{x}\pm i\partial _{y}\right) \right]
\end{eqnarray}%
we obtain the polar expressions%
\begin{equation}
a_{\pm }=\frac{1}{2}e^{\pm i\phi }\left( \rho +\frac{\partial }{\partial
\rho }\pm \frac{i}{\rho }\frac{\partial }{\partial \phi }\right) %
\mbox{\qquad}\bar{a}_{\pm }=\frac{1}{2}e^{\pm i\phi }\left[ \rho -\left( 
\frac{\partial }{\partial \rho }\pm \frac{i}{\rho }\frac{\partial }{\partial
\phi }\right) \right] .  \label{a_pm_expl}
\end{equation}%
Applying the annihilation operators to the ground state (\ref{G2}), we
recover (\ref{act_on_ground-1}) in the explicit form%
\begin{eqnarray}
a_{+}\psi _{0} &=&\frac{1}{2}\sqrt{\frac{1}{\pi \Gamma \left( s+1\right) }}%
\left( \rho +\allowbreak \left( s-\rho ^{2}\right) \frac{1}{\rho }-\frac{1}{%
\rho }s\right) e^{-\rho ^{2}/2}\rho ^{s}e^{i\phi \left( s+1\right) }=0
\label{a+0} \\
a_{-}\psi _{0} &=&\sqrt{s}\sqrt{\frac{1}{\pi \Gamma \left( s\right) }}%
e^{-\rho ^{2}/2}\left( \rho e^{i\phi }\right) ^{s-1},  \label{a-0}
\end{eqnarray}%
where the result in (\ref{a-0}) is formally equivalent to $\sqrt{s}\psi
_{0,-1}$ but as discussed above, does not correspond to any state in the
Fock space, and we treat as annihilation. For the general state (\ref%
{wavefunction}), using the notation of (\ref{sub-1}) $x=\rho ^{2}$, the $%
\rho $ derivative is%
\begin{equation}
\frac{\partial }{\partial \rho }\left[ e^{-\rho ^{2}/2}L_{n}^{m+s}\left(
\rho ^{2}\right) \left( \rho e^{i\phi }\right) ^{m+s}\right] =\frac{e^{-\rho
^{2}/2}\left( \rho e^{i\phi }\right) ^{m+s}}{\rho }\left[ -\rho ^{2}+m+s+2x%
\frac{d}{dx}\right] L_{n}^{m+s}\left( x\right)
\end{equation}%
providing%
\begin{equation}
\frac{\partial }{\partial \rho }\psi _{nm}=\left( -\rho +\frac{m+s}{\rho }%
\right) \psi _{nm}+\frac{2}{\rho }A_{nm}e^{-\rho ^{2}/2}\left( \rho e^{i\phi
}\right) ^{m+s}x\frac{d}{dx}L_{n}^{m+s}\left( x\right)
\end{equation}%
and the $\phi $ derivative is%
\begin{equation}
\frac{i}{\rho }\frac{\partial }{\partial \phi }\psi _{nm}=-\frac{m+s}{\rho }%
\psi _{nm}
\end{equation}%
so that%
\begin{eqnarray}
\left( \frac{\partial }{\partial \rho }+\frac{i}{\rho }\frac{\partial }{%
\partial \phi }\right) \psi _{nm} &=&-\rho \psi _{nm}+\frac{2}{\rho }%
A_{nm}e^{-\rho ^{2}/2}\left( \rho e^{i\phi }\right) ^{m+s}x\frac{d}{dx}%
L_{n}^{m+s}\left( x\right) \\
\left( \frac{\partial }{\partial \rho }-\frac{i}{\rho }\frac{\partial }{%
\partial \phi }\right) \psi _{nm} &=&\left( -\rho +2\frac{m+s}{\rho }\right)
\psi _{nm} \notag \\&& \qquad + \frac{2}{\rho }A_{nm}e^{-\rho ^{2}/2}\left( \rho e^{i\phi
}\right) ^{m+s}x\frac{d}{dx}L_{n}^{m+s}\left( x\right) .
\end{eqnarray}%
Then, using the identity \cite{Gradshteyn and Ryzhik} 
\begin{equation}
\frac{d}{dx}L_{a}^{b}\left( x\right) =-L_{a-1}^{b+1}\left( x\right)
\label{I1}
\end{equation}%
for the Laguerre functions, we calculate%
\begin{eqnarray}
a_{+}\psi _{nm}\left( \rho ,\phi \right) &=&\frac{1}{2}e^{i\phi }\left( \rho
+\frac{\partial }{\partial \rho }+\frac{i}{\rho }\frac{\partial }{\partial
\phi }\right) \psi _{nm} \\
&=&e^{i\phi }\frac{1}{\rho }A_{nm}e^{-\rho ^{2}/2}\left( \rho e^{i\phi
}\right) ^{m+s}x\frac{d}{dx}L_{n}^{m+s}\left( x\right) \\
&=&A_{nm}e^{-\rho ^{2}/2}\left( \rho e^{i\phi }\right) ^{m+s}e^{i\phi }\frac{%
\rho ^{2}}{\rho }\left( -L_{n-1}^{m+s+1}\right) ~,~n>0 \\
&=&-A_{nm}e^{-\rho ^{2}/2}\left( \rho e^{i\phi }\right)
^{m+s+1}L_{n-1}^{m+s+1}~,~n>0.
\end{eqnarray}%
Using (\ref{norm-2}) for $A_{nm}$ we obtain%
\begin{equation}
a_{+}\psi _{nm}=\left( -1\right) ^{n+1}\sqrt{\frac{\Gamma \left( n+1\right) 
}{\pi \Gamma \left( n+m+s+1\right) }}L_{n-1}^{m+s+1}\left( x\right) \left(
\rho e^{i\phi }\right) ^{m+s+1}e^{-\frac{\rho ^{2}}{2}}=\sqrt{n}\psi
_{n-1,m+1}  \label{a-exp}
\end{equation}%
as required by the first of (\ref{actions-1}). The second lowering operator
acts as%
\begin{eqnarray}
a_{-}\psi _{nm}\left( \rho ,\phi \right) &=&\frac{1}{2}e^{-i\phi }\left(
\rho +\frac{\partial }{\partial \rho }-\frac{i}{\rho }\frac{\partial }{%
\partial \phi }\right) \psi _{nm} \\
&=&\frac{m+s}{\rho }e^{-i\phi }\psi _{nm}+e^{-i\phi }\frac{1}{\rho }%
A_{nm}e^{-\rho ^{2}/2}\left( \rho e^{i\phi }\right) ^{m+s}x\frac{d}{dx}%
L_{n}^{m+s}\left( x\right)
\end{eqnarray}%
so the identities \cite{Gradshteyn and Ryzhik} 
\begin{eqnarray}
x\frac{d}{dx}L_{a}^{b}\left( x\right) &=&aL_{a}^{b}\left( x\right) -\left(
a+b\right) L_{a-1}^{b}\left( x\right)  \label{I2} \\
L_{a}^{b-1}\left( x\right) &=&L_{a}^{b}\left( x\right) -L_{a-1}^{b}\left(
x\right)
\end{eqnarray}%
lead to%
\begin{eqnarray}
a_{-}\psi _{nm}\left( \rho ,\phi \right) &=&A_{nm} e^{-\rho
^{2}/2} \left( \rho e^{i\phi }\right) ^{m+s-1}\Bigl[ \left( m+s\right) L_{n}^{m+s}\left( x\right) \notag \\
&&\mbox{\qquad}+ nL_{n}^{m+s}\left( x\right) -\left( n+m+s\right)
L_{n-1}^{m+s}\left( x\right)   \Bigr] \\
&=&\left( n+m+s\right) A_{nm}e^{-\rho ^{2}/2}\left( \rho e^{i\phi }\right)
^{m+s-1}\left( L_{n}^{m+s}\left( x\right) -L_{n-1}^{m+s}\left( x\right)
\right) \\
&=&\left( n+m+s\right) A_{nm}e^{-\rho ^{2}/2}\left( \rho e^{i\phi }\right)
^{m+s-1}L_{n}^{m+s-1}\left( x\right)
\end{eqnarray}%
providing%
\begin{eqnarray}
a_{-}\psi _{nm} &=&\left( -1\right) ^{n}\sqrt{\frac{\Gamma \left( n+1\right)
\left( m+s+n\right) ^{2}}{\pi \Gamma \left( n+m+s+1\right) }}e^{-\frac{\rho
^{2}}{2}}L_{n}^{m+s-1}\left( x\right) \left( \rho e^{i\phi }\right) ^{m+s-1}
\notag \\
&=&\sqrt{n+m+s}\psi _{n,m-1}  \label{a-b-xp}
\end{eqnarray}%
as required by the first of (\ref{actions-2}). The raising operator $\bar{a}%
_{+}$ acts as%
\begin{eqnarray}
\bar{a}_{+}\psi _{nm}\left( \rho ,\phi \right) &=&\frac{1}{2}e^{i\phi }\left[
\rho -\left( \frac{\partial }{\partial \rho }+\frac{i}{\rho }\frac{\partial 
}{\partial \phi }\right) \right] \psi _{nm}\left( \rho ,\phi \right) \\
&=&\rho e^{i\phi }\psi _{nm}-e^{i\phi }A_{nm}\frac{e^{-\rho ^{2}/2}\left(
\rho e^{i\phi }\right) ^{m+s}}{\rho }x\frac{d}{dx}L_{n}^{m+s}\left( x\right)
\end{eqnarray}%
so applying identity (\ref{I1}) and the identities \cite{Gradshteyn and
Ryzhik} 
\begin{equation}
L_{a}^{b}\left( x\right) =L_{a-1}^{b}\left( x\right) +L_{a}^{b-1}\left(
x\right)  \label{I3}
\end{equation}%
we calculate%
\begin{eqnarray}
\bar{a}_{+}\psi _{nm}\left( \rho ,\phi \right) &=&\rho e^{i\phi
}A_{nm}e^{-\rho ^{2}/2}\left( \rho e^{i\phi }\right) ^{m+s}\left(
L_{n}^{m+s}\left( x\right) -\frac{d}{dx}L_{n}^{m+s}\left( x\right) \right) \\
&=&A_{nm}e^{-\rho ^{2}/2}\left( \rho e^{i\phi }\right) ^{m+s+1}\left(
L_{n}^{m+s}\left( x\right) +L_{n-1}^{m+s+1}\left( x\right) \right) \\
&=&A_{nm}e^{-\rho ^{2}/2}\left( \rho e^{i\phi }\right)
^{m+s+1}L_{n}^{m+s+1}\left( x\right)
\end{eqnarray}%
providing%
\begin{eqnarray}
\bar{a}_{+}\psi _{nm} &=&\left( -1\right) ^{n}\sqrt{\frac{\Gamma \left(
n+1\right) }{\pi \Gamma \left( n+m+s+1\right) }}e^{-\rho ^{2}/2}\left( \rho
e^{i\phi }\right) ^{m+s+1}L_{n}^{m+s+1}\left( \rho ^{2}\right)  \\
&=&\sqrt{n+m+s+1}\psi _{n,m+1}
\end{eqnarray}%
confirming the second of (\ref{actions-1}). Finally, the action of the
second raising operator $\bar{a}_{-}$ is%
\begin{eqnarray}
\bar{a}_{-}\psi _{nm}\left( \rho ,\phi \right) &=&\frac{1}{2}e^{-i\phi }%
\left[ \rho -\left( \frac{\partial }{\partial \rho }-\frac{i}{\rho }\frac{%
\partial }{\partial \phi }\right) \right] \psi _{nm}\left( \rho ,\phi \right)
\\
&=&e^{-i\phi }\left[\left( \rho -\frac{m+s}{\rho }\right) \psi _{nm}-%
\frac{A_{nm}}{\rho }e^{-\rho ^{2}/2}\left( \rho e^{i\phi }\right) ^{m+s}x%
\frac{d}{dx}L_{n}^{m+s}\left( x\right)\right]
\end{eqnarray}%
so using the identity \cite{Gradshteyn and Ryzhik} 
\begin{equation}
x\frac{d}{dx}L_{a}^{b}\left( x\right) =\left( a+1\right) L_{a+1}^{b}\left(
x\right) -\left( a+b+1-x\right) L_{a}^{b}\left( x\right)
\end{equation}%
and (\ref{I1}) leads to%
\begin{eqnarray}
\bar{a}_{-}\psi _{nm}\left( \rho ,\phi \right) &=&-A_{nm}e^{-\rho
^{2}/2}\left( \rho e^{i\phi }\right) ^{m+s-1}\left[ \left( m+s-x\right)
L_{n}^{m+s}\left( x\right) +x\frac{d}{dx}L_{n}^{m+s}\left( x\right) \right]
\\
&=&-A_{nm}e^{-\rho ^{2}/2}\left( \rho e^{i\phi }\right) ^{m+s-1}\left[
\left( n+1\right) L_{n+1}^{m+s-1}\left( x\right) \right]
\end{eqnarray}%
we confirm%
\begin{eqnarray}
\bar{a}_{-}\psi _{nm} &=&-\left( -1\right) ^{n}\sqrt{\frac{\left( n+1\right)
^{2}\Gamma \left( n+1\right) }{\pi \Gamma \left( n+m+s+1\right) }}e^{-\rho
^{2}/2}L_{n+1}^{m+s-1}\left( \rho ^{2}\right) \left( \rho e^{i\phi }\right)
^{m+s-1}  \\
&=&\sqrt{n+1}\psi _{n+1,m-1}
\end{eqnarray}%
so that the solutions (\ref{wavefunction}) belong to the ladder
representation for any value of $s$.

Unlike the angular momentum $M=-i\partial _{\phi }$, which is diagonal in
polar coordinates, the remaining SU(2) generators are most conveniently
expressed in Cartesian coordinates%
\begin{eqnarray}
\Delta &=&\bar{a}_{+}a_{+}+\bar{a}_{-}a_{-}=N^{1}-N^{2}=\frac{1}{2}\left(
x^{2}-y^{2}-\frac{\partial ^{2}}{\partial x^{2}}+\frac{\partial ^{2}}{%
\partial y^{2}}\right) \\
Q &=&-i\left( \bar{a}_{+}a_{+}-\bar{a}_{-}a_{-}\right) =xy-\frac{\partial }{%
\partial x}\frac{\partial }{\partial y}
\end{eqnarray}%
from which it follows that%
\begin{eqnarray}
M\psi _{0} &=&s~\psi _{0}  \\
\Delta \psi _{0} &=&s\left[ \frac{x^{2}+y^{2}-s+1}{\left( x+iy\right) ^{2}}%
\right] \psi _{0}  \label{su-1} \\
Q\psi _{0} &=&is\left[ \frac{x^{2}+y^{2}-s+1}{\left( x+iy\right) ^{2}}\right]
\psi _{0}  
\end{eqnarray}%
and again we see that the SU(2) symmetry of the Hamiltonian is spontaneously
broken for $s\neq 0$.

Since the operator $\Delta $ is diagonal in Cartesian coordinates, its
action on the states is of special interest. From (\ref{gen-states}) and (%
\ref{Nab}) and the commutator 
\begin{equation}
\left[ \left( \bar{a}_{+}a_{+}+\bar{a}_{-}a_{-}\right) ,\left( \bar{a}%
_{+}\right) ^{\alpha }\left( \bar{a}_{-}\right) ^{\beta }\right] =\alpha 
\bar{a}_{+}^{\alpha -1}\left( \bar{a}_{-}\right) ^{\beta +1}+\beta \left( 
\bar{a}_{+}\right) ^{\alpha +1}\bar{a}_{-}^{\beta -1}
\end{equation}%
we obtain%
\begin{equation}
\Delta \zeta _{\alpha \beta }=\alpha \sqrt{\frac{\beta +1}{s+\alpha }}\zeta
_{\alpha -1,\beta +1}+\beta \sqrt{\frac{s+\alpha +1}{\beta }}\zeta _{\alpha
+1,\beta -1}+\left( \bar{a}_{+}\right) ^{\alpha }\left( \bar{a}_{-}\right)
^{\beta }\Delta \psi _{0}  \label{delta}
\end{equation}%
which cannot be diagonalized unless $s=0$ because (\ref{su-1}) shows that
the ground state is not an eigenstate of $\Delta $. Using (\ref{xi-psi}) to
construct the $s=0$ multiplets of for given $N$, it is easily shown, case by
case, that diagonalization of $\Delta $ recovers the standard Cartesian
description of the oscillator. Since $\Delta $ cannot be diagonalized on
states with $s\neq 0$, there is no unitary combination of the spherical
states $\psi _{nm}^{s\neq 0}$ equivalent to the familiar Cartesian states of
the harmonic oscillator.

\subsection{Number representation for $D=3$}

To obtain a number representation in $D=3$, we must simultaneously
diagonalize the operators $\mathbf{M}^{2}$ and $M$ expressed in terms of
creation/annihilation operators $\left( \bar{a}_{+},\overline{a}^{3},\bar{a}%
_{-}\right) $ for O(3) and $\left( \bar{a}_{+},\overline{a}^{0},\bar{a}%
_{-}\right) $ for O(2,1). As seen above, the states defined through the
actions of these operators on the ground states diagonalize $M$. However,
the Casimir operators%
\begin{equation}
\mathbf{M}^{2}=\frac{1}{2}M^{\mu \nu }M_{\mu \nu }=-\frac{1}{2}\left( \bar{a}%
^{\mu }a^{\nu }-\bar{a}^{\nu }a^{\mu }\right) \left( \bar{a}_{\mu }a_{\nu }-%
\bar{a}_{\nu }a_{\mu }\right) =N^{2}+N-\left( \bar{a}\cdot \bar{a}\right)
\left( a\cdot a\right)  \label{L-sq}
\end{equation}%
with total mode number 
\begin{equation}
N=\left\{ 
\begin{array}{l}
\overline{a}^{1}a^{1}+\overline{a}^{2}a^{2}+\overline{a}^{3}a^{3}=\bar{a}%
_{+}a_{-}+\bar{a}_{-}a_{+}+\bar{a}^{3}a^{3}\;,\mbox{\qquad}\text{O(3)} \\%
[0.3cm] 
\overline{a}^{1}a^{1}+\overline{a}^{2}a^{2}+\overline{a}^{0}a^{0}=\bar{a}%
_{+}a_{-}+\bar{a}_{-}a_{+}-\overline{a}^{0}a^{0}\;,\mbox{\qquad}\text{O(2,1)}%
\end{array}%
\right.  \label{N-3d}
\end{equation}%
and scalar products%
\begin{equation}
\bar{a}\cdot \bar{a}=\left\{ 
\begin{array}{l}
2\bar{a}_{+}\bar{a}_{-}+\bar{a}^{3}\bar{a}^{3}\;,\mbox{\qquad}\text{O(3)} \\%
[0.3cm] 
2\bar{a}_{+}\bar{a}_{-}-\bar{a}^{0}\bar{a}^{0}\;,\mbox{\qquad}\text{O(2,1)}%
\end{array}%
\right.  \label{aa-3d}
\end{equation}%
\begin{equation}
a\cdot a=\left\{ 
\begin{array}{l}
2a_{+}a_{-}+a^{3}a^{3}\;,\mbox{\qquad}\text{O(3)} \\[0.3cm] 
2a_{+}a-a^{0}a^{0}\;,\mbox{\qquad}\text{O(2,1)}%
\end{array}%
\right.  \label{aa-3d-2}
\end{equation}%
remain non-diagonal. Nevertheless, $\mathbf{M}^{2}$ must be block diagonal
with respect to $N$ and $M$ and so studying the expected multiplicity of the
mass/energy states leads to a characterization of the oscillator states.
Recall that despite the sign of the term $-\overline{a}^{0}a^{0}$ in the
second of (\ref{N-3d}), the total mass/energy of the O(2,1) oscillator was
found in equation (\ref{energy-3d})
to be positive definite --- because solutions
(\ref{solu-4}) and (\ref{solu-5}) are of the Feynman, Kislinger, and Ravndal type
for which (\ref {a0}) requires $n^{0}\leq 0$, the timelike
modes contribute positive mass/energy. We therefore expect that the O(3) and O(2,1)
will have similar multiplicity structure, which is verified by examining the
wavefunctions as representations of the respective symmetry groups.

The O(3) wavefunctions (\ref{solu-2}) for $s=0$ depend on $\theta $ and $%
\phi $ through the spherical harmonics%
\begin{equation}
Y_{l}^{m}\left( \theta ,\phi \right) =C_{lm}P_{l}^{m}\left( \cos \theta
\right) e^{im\phi }
\end{equation}%
and thus provide the familiar $\left( 2l+1\right) $-dimensional
representation of O(3) through 
\begin{equation}
\mathbf{L}^{2}Y_{l}^{m}\left( \theta ,\phi \right) =l\left( l+1\right)
Y_{l}^{m}\left( \theta ,\phi \right) \mbox{\qquad}L^{3}Y_{l}^{m}\left(
\theta ,\phi \right) =mY_{l}^{m}\left( \theta ,\phi \right)
\end{equation}%
\begin{equation}
L^{\pm }Y_{l}^{m}\left( \theta ,\phi \right) =\sqrt{\left( l\mp m\right)
\left( l\pm m+1\right) }Y_{l}^{m}\left( \theta ,\phi \right)  \label{L+-}
\end{equation}%
with allowed values%
\begin{equation}
l=0,1,...\mbox{\qquad}m=-l,-l+1,...,l-1,l.
\end{equation}%
Similarly, the O(2,1) wavefunctions (\ref{solu-4}) for $s=0$ depend on $%
\beta $ and $\phi $ through the functions%
\begin{equation}
\hat{Y}_{l}^{m}\left( \beta ,\phi \right) =C_{lm}\hat{P}_{l}^{m}\left( \sinh
\beta \right) e^{im\phi }
\end{equation}%
and it follows from (\ref{beta}) and (\ref{sub-4}) that%
\begin{equation}
\mathbf{\Lambda }^{2}\psi _{nlm}^{\text{O(2,1)},s=0}=l\left( l+1\right) \psi
_{nlm}^{\text{O(2,1)},s=0}\mbox{\qquad}M\psi _{nlm}^{\text{O(2,1)}%
,s=0}=m\psi _{nlm}^{\text{O(2,1)},s=0}.  \label{100}
\end{equation}%
The remaining O(2,1) boost generators are%
\begin{equation}
A^{\pm }=A^{1}\pm iA^{2}=-ie^{\pm i\phi }\left( \partial _{\beta }\pm i\frac{%
\sinh \beta }{\cosh \beta }\partial _{\phi }\right)
\end{equation}%
and for $s=0$, where we set $\zeta =\sinh \beta $, they take the form%
\begin{equation}
A^{\pm }=-i\frac{e^{\pm i\phi }}{\left( 1+\zeta ^{2}\right) ^{1/2}}\left[
\left( 1+\zeta ^{2}\right) \partial _{\zeta }\pm i\zeta \partial _{\phi }%
\right]
\end{equation}%
and%
\begin{equation}
A^{\pm }\hat{Y}_{l}^{m}\left( \beta ,\phi \right) =-iC_{lm}\frac{e^{i\left(
m\pm 1\right) \phi }}{\left( 1+\zeta ^{2}\right) ^{1/2}}\left[ \left(
1+\zeta ^{2}\right) \partial _{\zeta }\hat{P}_{l}^{m}\left( \zeta \right)
\mp m\zeta \hat{P}_{l}^{m}\left( \zeta \right) \right] .
\end{equation}%
Using the identities \cite{Gradshteyn and Ryzhik}%
\begin{eqnarray}
\left( 1+\zeta ^{2}\right) \frac{d}{d\zeta }\hat{P}_{\nu }^{\mu }\left(
\zeta \right) &=&\sqrt{1+\zeta ^{2}}\hat{P}_{\nu }^{\mu +1}\left( \zeta
\right) +\mu \zeta \hat{P}_{\nu }^{\mu }\left( \zeta \right) \\
\left( 1+\zeta ^{2}\right) \frac{d}{d\zeta }\hat{P}_{\nu }^{\mu }\left(
\zeta \right) &=&\left( \nu -\mu +1\right) \left( \nu +\mu \right) \sqrt{%
1+\zeta ^{2}}\hat{P}_{\nu }^{\mu -1}\left( \zeta \right) -\mu \zeta \hat{P}%
_{\nu }^{\mu }\left( \zeta \right)
\end{eqnarray}%
we find%
\begin{eqnarray}
A^{+}\hat{Y}_{l}^{m}\left( \beta ,\phi \right) &=&-iC_{lm}\frac{e^{i\left(
m+1\right) \phi }}{\left( 1+\zeta ^{2}\right) ^{1/2}}\left( \sqrt{1+\zeta
^{2}}\hat{P}_{l}^{m+1}\left( \zeta \right) +m\zeta \hat{P}_{l}^{m}\left(
\zeta \right) -m\zeta \hat{P}_{l}^{m}\left( \zeta \right) \right) \\
&=&-iC_{lm}e^{i\left( m+1\right) \phi }\hat{P}_{l}^{m+1}\left( \zeta \right)
\end{eqnarray}%
\begin{eqnarray}
A^{-}\hat{Y}_{l}^{m}\left( \beta ,\phi \right) &=&-iC_{lm}\frac{e^{i\left(
m-1\right) \phi }}{\left( 1+\zeta ^{2}\right) ^{1/2}}%
\Bigl%
(\left( l-m+1\right) \left( l+m\right) \sqrt{1+z^{2}}\hat{P}_{l}^{m-1}\left(
\zeta \right) -m\zeta \hat{P}_{l}^{m}\left( \zeta \right) \notag \\
&&\mbox{\qquad}\mbox{\qquad}\mbox{\qquad}+m\zeta \hat{P}_{l}^{m}\left( \zeta \right) 
\Bigr%
) \\
&=&-i\left( l-m+1\right) \left( l+m\right) C_{lm}e^{i\left( m-1\right) \phi }%
\hat{P}_{l}^{m-1}\left( \zeta \right)
\end{eqnarray}%
and so that the boost operators $A^{\pm }$ raise and lower the $m$
eigenvalue as%
\begin{equation}
A^{\pm }\hat{Y}_{l}^{m}\left( \beta ,\phi \right) =\sqrt{\left( l\mp
m\right) \left( l\pm m+1\right) }\hat{Y}_{l}^{m\pm 1}\left( \beta ,\phi
\right)  \label{101}
\end{equation}%
comparable to the action of $L^{\pm }$ in (\ref{L+-}). It follows from (\ref%
{100}) and (\ref{101}) that the hyperangular functions (\ref{solu-4})
provide a $\left( 2l+1\right) $-dimensional representation of O(2,1) with
the same multiplicity structure found in the $s=0$ solutions for O(3). Since
the unitary representations of the non-compact group O(2,1) should be
infinite-dimensional, the $s=0$ solutions appear to violate unitarity.

The $s=1/2$ wavefunctions (\ref{wf-half}) for O(3) depend on $\theta $ and $%
\phi $ through the angular functions%
\begin{equation}
\hat{\chi}_{m}^{l}\left( \theta ,\phi \right) =F_{m}^{l}\left( z\right)
e^{i\left( m+1/2\right) \phi }=C_{lm}\left( 1+z^{2}\right) ^{\frac{1}{4}}%
\hat{P}_{m}^{l}\left( z\right) e^{i\left( m+1/2\right) \phi }  \label{ang_hf}
\end{equation}%
where $z=\cot \theta $, and it follows from (\ref{ang-2}), (\ref{sub-5}) and
(\ref{sub-6}) that%
\begin{equation}
\mathbf{L}^{2}\hat{\chi}_{m}^{l}\left( \theta ,\phi \right) =\left(
l^{2}-1/4\right) \hat{\chi}_{m}^{l}\left( \theta ,\phi \right) \mbox{\qquad}M%
\hat{\chi}_{m}^{l}\left( \theta ,\phi \right) =\left( m+1/2\right) \hat{\chi}%
_{m}^{l}\left( \theta ,\phi \right) .
\end{equation}%
Similarly, the $s=1/2$ wavefunctions (\ref{solu-5}) for O(2,1) depend on $%
\beta $ and $\phi $ through the hyperangular functions%
\begin{equation}
\chi _{m}^{l}\left( \beta ,\phi \right) =G_{m}^{l}\left( \zeta \right)
e^{i\left( m+1/2\right) \phi }=C_{lm}\left( 1-\zeta ^{2}\right) ^{\frac{1}{4}%
}P_{m}^{l}\left( \zeta \right) e^{i\left( m+1/2\right) \phi }
\end{equation}%
where $\zeta =\tanh \beta $ and 
\begin{equation}
\mathbf{\Lambda }^{2}\chi _{m}^{l}\left( \beta ,\phi \right) =\left(
l^{2}-1/4\right) \chi _{m}^{l}\left( \beta ,\phi \right) \mbox{\qquad}M\chi
_{m}^{l}\left( \beta ,\phi \right) =\left( m+1/2\right) \chi _{m}^{l}\left(
\beta ,\phi \right) .
\end{equation}%
In terms of the parameters (\ref{subs_hf}) and (\ref{subs_hf2}) for $s=1/2$,
the non-diagonal operators for O(3) and O(2,1) take the forms%
\begin{equation}
L^{\pm }=L^{1}\pm iL^{2}=e^{\pm i\phi }\left[ \pm \left( 1+z^{2}\right)
\partial _{z}-iz\partial _{\phi }\right] \mbox{\qquad}\text{O(3)}
\end{equation}%
\begin{equation}
A^{\pm }=A^{1}\pm iA^{2}=e^{\pm i\phi }\left[ -i\left( 1-\zeta ^{2}\right)
\partial _{\zeta }\pm \zeta \partial _{\phi }\right] \mbox{\qquad}\text{%
O(2,1).}
\end{equation}%
For O(3)%
\begin{equation}
L^{\pm }\hat{\chi}_{m}^{l}\left( \theta ,\phi \right) =C_{lm}e^{\pm i\phi }%
\left[ \pm \left( 1+z^{2}\right) \partial _{z}+\left( m+1/2\right) z\right]
\left( 1+z^{2}\right) ^{\frac{1}{4}}\hat{P}_{m}^{l}\left( z\right)
e^{i\left( m+1/2\right) \phi }
\end{equation}%
where%
\begin{equation}
\left( 1+z^{2}\right) \partial _{z}\left( 1+z^{2}\right) ^{\frac{1}{4}}\hat{P%
}_{m}^{l}\left( z\right) =\left( 1+z^{2}\right) ^{\frac{1}{4}}\left[ \frac{1%
}{2}z\hat{P}_{m}^{l}\left( z\right) +\left( 1+z^{2}\right) \frac{d}{dz}\hat{P%
}_{m}^{l}\left( z\right) \right]
\end{equation}%
so that 
\begin{equation}
L^{\pm }\hat{\chi}_{m}^{l}\left( \theta ,\phi \right) =C_{lm}e^{i\left(
m+1/2\pm 1\right) \phi }\left( 1+z^{2}\right) ^{\frac{1}{4}}\left[ \pm
\left( 1+z^{2}\right) \frac{d}{dz}+\left( m+1/2\pm 1/2\right) z\right] \hat{P%
}_{m}^{l}\left( z\right)
\end{equation}%
and using the identities \cite{Gradshteyn and Ryzhik}%
\begin{equation}
\left( 1+z^{2}\right) \frac{d}{dz}\hat{P}_{\nu }^{\mu }=-\left( \mu -\nu
-1\right) \hat{P}_{\nu +1}^{\mu }-\left( \nu +1\right) z\hat{P}_{\nu }^{\mu
}=\left( \mu +\nu \right) \hat{P}_{\nu -1}^{\mu }+\nu z\hat{P}_{\nu }^{\mu }
\end{equation}%
one is led to%
\begin{eqnarray}
L^{+}\hat{\chi}_{m}^{l}\left( \theta ,\phi \right) &=&C_{lm}e^{i\left(
m+1/2+1\right) \phi }\left( 1+z^{2}\right) ^{\frac{1}{4}}%
\Bigl%
[-\left( l-m-1\right) \hat{P}_{m+1}^{l}-\left( m+1\right) z\hat{P}_{m}^{l} \notag \\
&&\mbox{\qquad}\mbox{\qquad}+\left( m+1\right) z\hat{P}_{m}^{l}\left(
z\right) 
\Bigr%
] \\
&=&-\left( l-m-1\right) C_{lm}e^{i\left( m+1/2+1\right) \phi }\left(
1+z^{2}\right) ^{\frac{1}{4}}\hat{P}_{m+1}^{l}
\end{eqnarray}%
and%
\begin{eqnarray}
L^{-}\hat{\chi}_{m}^{l}\left( \theta ,\phi \right) &=&C_{lm}e^{i\left(
m+1/2-1\right) \phi }\left( 1+z^{2}\right) ^{\frac{1}{4}}\Bigl[ -\left(
\left( l+m\right) \hat{P}_{m-1}^{l}+mz\hat{P}_{m}^{l}\right) \notag \\
&&\mbox{\qquad}+mz\hat{P}%
_{m}^{l}\left( z\right) \Bigr] \\
&=&-\left( l+m\right) C_{lm}e^{i\left( m+1/2-1\right) \phi }\left(
1+z^{2}\right) ^{\frac{1}{4}}\hat{P}_{m-1}^{l}.
\end{eqnarray}%
The actions of $L^{\pm }$ and a similar calculation for $A^{\pm }$ using the
identities \cite{Gradshteyn and Ryzhik} 
\begin{equation}
\left( 1-\zeta ^{2}\right) \frac{d}{d\zeta }P_{\nu }^{\mu }=\left( \mu -\nu
-1\right) P_{\nu +1}^{\mu }+\left( \nu +1\right) \zeta P_{\nu }^{\mu
}=\left( \mu +\nu \right) P_{\nu -1}^{\mu }-\nu zP_{\nu }^{\mu }
\end{equation}%
leads to 
\begin{equation}
L^{\pm }\hat{\chi}_{m}^{l}\left( \theta ,\phi \right) =\hat{c}\left(
l,m\right) \hat{\chi}_{m\pm 1}^{l}\left( \theta ,\phi \right) \mbox{\qquad}%
A^{\pm }\chi _{m}^{l}\left( \beta ,\phi \right) =c\left( l,m\right) \chi
_{m\pm 1}^{l}\left( \beta ,\phi \right)  \label{inf-rep}
\end{equation}%
where $\hat{c}\left( l,m\right) $ and $c\left( l,m\right) $ are combinations
of the eigenvalues. Since $L^{\pm }$ and $A^{\pm }$ act on the lower index
(the associated Legendre functions $P_{\nu }^{\mu }$ and $\hat{P}_{\nu
}^{\mu }$ are nonzero for $\nu \geq 0$ and $\nu \geq \left\vert \mu
\right\vert $), there is no upper bound on the action of the raising
operators $L^{+}$ and $A^{+}$, but from 
\begin{equation}
\hat{P}_{n}^{n}=\frac{\left( 2n\right) !}{2^{n}n!}\left( 1+z^{2}\right)
^{n/2}\mbox{\qquad}P_{n}^{n}\left( \zeta \right) =\left( -1\right) ^{n}\frac{%
\left( 2n\right) !}{2^{n}n!}\left( 1+\zeta ^{2}\right) ^{n/2}
\end{equation}%
we find the lower bounds 
\begin{equation}
L^{-}\hat{\chi}_{m}^{m}\left( \theta ,\phi \right) =A^{-}\chi _{m}^{m}\left(
\beta ,\phi \right) =0.
\end{equation}%
The functions $F$ and $G$ therefore provide infinite-dimensional
representations of O(3) and O(2,1), leading to mass/energy states of
infinite degeneracy, appropriate to the non-compact O(2,1) but apparently
violating unitarity for O(3).

Since the multiplicity structure of the wavefunctions (\ref{solu-2}) to (\ref%
{solu-5}) depends on $s$ but not on the relevant symmetry group, we study
their eigenvalue content together. We know that for the standard Cartesian
states,%
\begin{equation}
\left[ M,N^{1}\right] \neq 0\mbox{\qquad}\left[ M,N^{2}\right] \neq 0%
\mbox{\qquad}\left[ N,N^{\parallel }\right] =\left[ M,N^{\parallel }\right]
=0
\end{equation}%
where the longitudinal component, relative to the choice of $x-y$ plane as
locus of observable angular momentum, is%
\begin{equation}
N^{\parallel }=\left\{ 
\begin{array}{lll}
N^{3} &  & \text{O(3)} \\[0.2cm] 
N^{0} &  & \text{O(2,1)}%
\end{array}%
\right. .
\end{equation}%
Therefore, the matrix representation of $\mathbf{M}^{2}$ reduces to coherent
subspaces labeled by eigenvalues $N$ and $n^{\parallel }$, and a convenient
parameterization of Cartesian states is%
\begin{equation}
\left( 
\begin{array}{l}
n^{1} \\ 
n^{2} \\ 
n^{\parallel }%
\end{array}%
\right) =\left( 
\begin{array}{c}
k \\ 
N-n^{\parallel }-k \\ 
n^{\parallel }%
\end{array}%
\right)  \label{C-param}
\end{equation}%
with%
\begin{equation}
n^{\parallel }=0,1,...,N,\mbox{\qquad}k=0,1,...,\left( N-n^{\parallel
}\right) .
\end{equation}%
The number of states for given $N$ and $n^{\parallel }$ is therefore $%
N-n^{\parallel }+1$, and the total number of states with mode number $N$ is%
\begin{equation}
\dsum\limits_{n^{\parallel }=0}^{N}\left( N+1-n^{\parallel }\right) =\left(
N+1\right) \left( N+1\right) -\frac{N\left( N+1\right) }{2}=\frac{\left(
N+1\right) \left( N+2\right) }{2}\ .  \label{multi}
\end{equation}%
For $s=0$, we extend (\ref{gen-states}) and construct excited states through 
\begin{equation}
\zeta _{\alpha \beta \gamma }=\frac{1}{\sqrt{\alpha !\beta !\gamma !}}\left( 
\bar{a}_{+}\right) ^{\alpha }\left( \bar{a}_{-}\right) ^{\beta }\left( \bar{a%
}^{\parallel }\right) ^{\gamma }\psi _{0}  \label{gen-states-3d}
\end{equation}%
which are eigenstates of $N$ and $M$ with 
\begin{equation}
N\zeta _{\alpha \beta \gamma }=\left( \alpha +\beta +\gamma \right) \zeta
_{\alpha \beta \gamma }\mbox{\qquad\qquad}M\zeta _{\alpha \beta \gamma
}=\left( \alpha -\beta \right) \zeta _{\alpha \beta \gamma }  \label{N-m}
\end{equation}%
so that the states $\zeta _{\alpha \beta \gamma }$ are precisely the states
found by diagonalizing $M$ in the Cartesian picture. Acting on (\ref%
{gen-states-3d}) with (\ref{L-sq}) leads to%
\begin{eqnarray}
\mathbf{M}^{2}\zeta _{\alpha \beta \gamma } &=&\left[ N\left( N+1\right)
-4\alpha \beta -\gamma \left( \gamma -1\right) \right] \zeta _{\alpha \beta
\gamma }  \notag \\
&&\mbox{\qquad\qquad}-2\sqrt{\left( \alpha +1\right) \left( \beta +1\right)
\gamma \left( \gamma -1\right) }\zeta _{\left( \alpha +1\right) \left( \beta
+1\right) \left( \gamma -2\right) }  \notag \\
&&\mbox{\qquad\qquad}-2\sqrt{\alpha \beta \left( \gamma +2\right) \left(
\gamma +1\right) }\zeta _{\left( \alpha -1\right) \left( \beta -1\right)
\left( \gamma +2\right) }  \label{L-sq-act}
\end{eqnarray}%
so that the states $\zeta _{\alpha \beta \gamma }$ are not generally
eigenstates of $\mathbf{M}^{2}$, but as expected are mixtures of states with 
$\left( \alpha \pm 1,\beta \pm 1,\gamma \mp 2\right) $ and fixed $M$
eigenvalue%
\begin{equation}
m=\left( \alpha \pm 1\right) -\left( \beta \pm 1\right) =\alpha -\beta .
\end{equation}%
It follows from (\ref{L-sq-act}) that 
\begin{equation}
\mathbf{M}^{2}\zeta _{N00}=N\left( N+1\right) \zeta _{N00}\mbox{\qquad}%
M\zeta _{N00}=N\zeta _{N00}
\end{equation}%
\begin{equation}
\mathbf{M}^{2}\zeta _{0N0}=N\left( N+1\right) \zeta _{0N0}\mbox{\qquad}%
M\zeta _{0N0}=-N\zeta _{0N0}
\end{equation}%
and so the allowed eigenvalues of $M$ 
\begin{equation}
m=\alpha -\beta =-l,-l+1,...,l-1,l
\end{equation}%
are consistent with the parameter range 
\begin{equation}
\alpha ,\beta =0,1,...,N.
\end{equation}%
Generally, as demonstrated in \cite{talk} by exploiting the invariance of tr$%
\left( \mathbf{M}^{2}\right) $ under unitary transformations, the Casimir
content of the states $\zeta _{\alpha \beta \gamma }$ is%
\begin{equation}
l=N,N-2,...,N-\func{int}\left( N/2\right)  \label{l}
\end{equation}%
and since the multiplicity of $l$-states is $2l+1$, the multiplicity of
states with total mode number $N$ is 
\begin{equation}
\sum_{k=0}^{\func{int}\left( N/2\right) }2\left( N-2k\right) +1=\frac{\left(
N+1\right) \left( N+2\right) }{2}
\end{equation}%
in agreement with (\ref{multi}). Since diagonalization of $\mathbf{M}^{2}$
does not mix states of different $m$, states $\psi _{lm}$ have mode number $%
N $ that depends on $l$, with Casimir eigenvalues given in (\ref{l}), but
not on $m$, so there must be a principal quantum number $n$ that complements
the contribution of $l$ to energy, incrementing by 2 when $l$ is decremented
by 1. Thus, the mode number can be written%
\begin{equation}
N=2n+l,\mbox{\qquad}n=0,1,2,...,N
\end{equation}%
and the total energy must be%
\begin{equation}
E=\omega \left( 2n+l+\frac{3}{2}\right)
\end{equation}%
in agreement with the solution (\ref{energy-3d}) to the Schrodinger equation.

According to (\ref{gen-states-3d}) and (\ref{N-m}) the $N=1$ states
constitute the $l=1$ vector multiplet%
\begin{equation}
\zeta ^{\left( 1\right) }=\left( 
\begin{array}{c}
\zeta _{001} \\ 
\zeta _{010} \\ 
-\zeta _{100}%
\end{array}%
\right) =\left( 
\begin{array}{c}
\bar{a}_{-} \\ 
\bar{a}^{\parallel } \\ 
-\bar{a}_{+}%
\end{array}%
\right) \psi _{0}^{s=0},
\end{equation}%
which we order according to the eigenvalues $m=-1,0,1$ found by
diagonalizing $M$ on the $N=1$ multiplet of Cartesian states%
\begin{equation}
\varphi ^{\left( 1\right) }=\left( 
\begin{array}{l}
\varphi _{100} \\ 
\varphi _{010} \\ 
\varphi _{001}%
\end{array}%
\right) =\left( 
\begin{array}{l}
\bar{a}_{1} \\ 
\bar{a}_{2} \\ 
\bar{a}^{\parallel }%
\end{array}%
\right) \varphi _{0}.
\end{equation}%
Applying the creation/annihilation operators in polar parameterizations (\ref%
{pol-coord}) 
\begin{equation}
\bar{a}_{\pm }=\frac{1}{2}e^{\pm i\phi }\left( \rho \sin \theta -\sin \theta
\partial _{\rho }-\frac{\cos \theta }{\rho }\partial _{\theta }\mp \frac{i}{%
\rho \sin \theta }\partial _{\phi }\right)
\end{equation}%
\begin{equation}
\overline{a}^{3}=\frac{1}{\sqrt{2}}\left( \rho \cos \theta -\cos \theta
\partial _{\rho }+\frac{\sin \theta }{\rho }\partial _{\beta }\right)
\end{equation}%
for O(3) and%
\begin{equation}
\bar{a}_{\pm }=\frac{1}{2}e^{\pm i\phi }\left( \rho \cosh \beta -\cosh \beta
\partial _{\rho }+\frac{\sinh \beta }{\rho }\partial _{\beta }\mp \frac{i}{%
\rho \cosh \beta }\partial _{\phi }\right)
\end{equation}%
\begin{equation}
\bar{a}^{0}=\frac{1}{\sqrt{2}}\left( \rho \sinh \beta -\sinh \beta \partial
_{\rho }+\frac{\cosh \beta }{\rho }\partial _{\beta }\right)
\end{equation}%
for O(2,1) to the ground states, 
\begin{equation}
\psi _{0}^{\text{O(3)},s=0}=\psi _{0}^{\text{O(2,1)},s=0}=A_{0}e^{-\rho
^{2}/2}
\end{equation}%
we obtain%
\begin{equation}
\zeta ^{\left( 1\right) }=A_{0}\rho \left( 
\begin{array}{c}
\sin \theta e^{-i\phi } \\ 
\sqrt{2}\cos \theta \\ 
-\sin \theta e^{i\phi }%
\end{array}%
\right) e^{-\rho ^{2}/2}\mbox{\qquad}\text{O(3)}  \label{110}
\end{equation}%
\begin{equation}
\zeta ^{\left( 1\right) }=A_{0}\rho \left( 
\begin{array}{c}
\cosh \beta e^{-i\phi } \\ 
\sqrt{2}\sinh \beta \\ 
-\cosh \beta e^{i\phi }%
\end{array}%
\right) e^{-\rho ^{2}/2}\mbox{\qquad}\text{O(2,1)}.  \label{120}
\end{equation}%
Wavefunctions (\ref{110}) and (\ref{120}) are seen to agree with the $l=1$
vector multiplet found from (\ref{solu-2}) and (\ref{solu-4}) using%
\begin{equation}
P_{1}^{1}\left( z\right) =-\sqrt{1-z^{2}}\mbox{\qquad}P_{1}^{0}\left(
z\right) =z\mbox{\qquad}P_{1}^{-1}\left( z\right) =\sqrt{1-z^{2}}
\end{equation}%
\begin{equation}
\hat{P}_{1}^{1}\left( \zeta \right) =-\sqrt{1+\zeta ^{2}}\mbox{\qquad}\hat{P}%
_{1}^{0}\left( \zeta \right) =\zeta \mbox{\qquad}\hat{P}_{1}^{-1}\left(
\zeta \right) =\sqrt{1+\zeta ^{2}}.
\end{equation}%
The $l=1$ multiplet of the spherical harmonics $Y_{l}^{m}\left( \theta ,\phi
\right) $ and $\hat{Y}_{l}^{m}\left( \beta ,\phi \right) $ have the
well-known property that the three components form a unit vector, so%
\begin{equation}
\rho \left( 
\begin{array}{c}
Y_{1}^{-1} \\ 
Y_{1}^{0} \\ 
Y_{1}^{1}%
\end{array}%
\right) =\left( 
\begin{array}{c}
x_{-} \\ 
x^{3} \\ 
-x_{+}%
\end{array}%
\right) =\frac{1}{\sqrt{2}}\left( 
\begin{array}{c}
x-iy \\ 
\sqrt{2}z \\ 
-x-iy%
\end{array}%
\right) \mbox{\qquad}\text{O(3)}
\end{equation}%
\begin{equation}
\rho \left( 
\begin{array}{c}
\hat{Y}_{1}^{-1} \\ 
\hat{Y}_{1}^{0} \\ 
\hat{Y}_{1}^{1}%
\end{array}%
\right) =\left( 
\begin{array}{c}
x_{-} \\ 
x^{0} \\ 
-x_{+}%
\end{array}%
\right) =\frac{1}{\sqrt{2}}\left( 
\begin{array}{c}
x-iy \\ 
\sqrt{2}t \\ 
-x-iy%
\end{array}%
\right) \mbox{\qquad}\text{O(2,1)}
\end{equation}%
in the basis that diagonalizes the $3\times 3$ matrix representation of $M$,
which may be verified using the parameterizations (\ref{pol-coord}).

The first level of excited states was found by acting on the ground state
with the operator multiplet $\left( \bar{a}_{-},\bar{a}^{\parallel },-\bar{a}%
_{+}\right) $ which we regard as the fundamental representation of a set of
irreducible tensor operators constructed successively by taking irreducible
tensor products%
\begin{equation}
\bar{a}_{m}^{\left( j\pm 1\right) }=\sum_{m_{2}=-1,0,1}\left\langle
j~m-m_{2}~1~m_{2}~|~j~1~j\pm 1~m\right\rangle \bar{a}_{m-m_{2}}^{\left(
j\right) }\bar{a}_{m_{2}}^{\left( 1\right) }
\end{equation}%
where $\bar{a}_{m}^{\left( j\pm 1\right) }$ is an irreducible tensor
operator of rank $j\pm 1$, $\bar{a}_{m_{2}}^{\left( 1\right) }$ is the
vector operator, and $\left\langle j~m-m_{2}~1~m_{2}~|~j~1~j\pm
1~m\right\rangle $ is the appropriate Clebsch-Gordan coefficient. Thus,
according to (\ref{multi}), the $N=2$ states have total multiplicity of $6$,
which by (\ref{l}) must include the five $l=2$ states and the $l=0$ singlet
state. The two irreducible tensor operators that can be constructed from the
vector operator are the singlet ($l=m=0$)%
\begin{equation}
\bar{a}_{0}^{\left( 0\right) }=-\frac{1}{\sqrt{3}}\left( 2\bar{a}_{+}\bar{a}%
_{-}+\bar{a}_{3}\bar{a}_{3}\right) =-\frac{1}{\sqrt{3}}\bar{a}\cdot \bar{a}
\end{equation}%
and the $l=2$ operators%
\begin{eqnarray}
\bar{a}_{-2}^{\left( 2\right) } &=&\bar{a}_{-1}^{\left( 1\right) }\bar{a}%
_{-1}^{\left( 1\right) }=\left( \bar{a}_{-}\right) ^{2} \\
\bar{a}_{-1}^{\left( 2\right) } &=&\frac{1}{\sqrt{2}}\left( \bar{a}%
_{0}^{\left( 1\right) }\bar{a}_{-1}^{\left( 1\right) }+\bar{a}_{-1}^{\left(
1\right) }\bar{a}_{0}^{\left( 1\right) }\right) =-\sqrt{2}\bar{a}_{-}\bar{a}%
_{3} \\
\bar{a}_{0}^{\left( 2\right) } &=&\frac{1}{\sqrt{6}}\left( \bar{a}%
_{1}^{\left( 1\right) }\bar{a}_{-1}^{\left( 1\right) }+2\bar{a}_{0}^{\left(
1\right) }\bar{a}_{0}^{\left( 1\right) }+\bar{a}_{-1}^{\left( 1\right) }\bar{%
a}_{1}^{\left( 1\right) }\right) =-\frac{2}{\sqrt{6}}\left( \bar{a}_{+}\bar{a%
}_{-}-\bar{a}_{3}\bar{a}_{3}\right) \\
\bar{a}_{1}^{\left( 2\right) } &=&\frac{1}{\sqrt{2}}\left( \bar{a}%
_{1}^{\left( 1\right) }\bar{a}_{0}^{\left( 1\right) }+\bar{a}_{1-1}^{\left(
1\right) }\bar{a}_{1}^{\left( 1\right) }\right) =\sqrt{2}\bar{a}_{+}\bar{a}%
_{3} \\
\bar{a}_{2}^{\left( 2\right) } &=&\bar{a}_{1}^{\left( 1\right) }\bar{a}%
_{1}^{\left( 1\right) }=\left( \bar{a}_{+}\right) ^{2}
\end{eqnarray}%
which are precisely the operators found by diagonalizing the matrix
representation of $\mathbf{L}^{2}$. In this way, the complete set of
spherical polar harmonic oscillators in 3 dimensions can be constructed from
the ground state.

For the $s=1/2$ wave functions, the attempt to build excited states from the
ground state and the operator multiplet $\left( \bar{a}_{-},\bar{a}%
^{\parallel },-\bar{a}_{+}\right) $ fails immediately. The infinitely
degenerate ground states found from (\ref{wf-half}) and (\ref{solu-5}) are%
\begin{equation}
\psi _{0}=Ae^{-\rho ^{2}/2}\rho ^{-1/2}\left( 1+z^{2}\right) ^{\frac{1}{4}}%
\hat{P}_{m}^{0}\left( z\right) e^{i\left( m+\frac{1}{2}\right)
},\;\;m=0,1,2,....,\mbox{\qquad}\text{O(3)}
\end{equation}%
\begin{equation}
\psi _{0}=Ae^{-\rho ^{2}/2}\rho ^{-1/2}\left( 1-\zeta ^{2}\right) ^{\frac{1}{%
4}}P_{m}^{0}\left( \zeta \right) e^{i\left( m+\frac{1}{2}\right)
},\;\;m=0,1,2,....,\mbox{\qquad}\text{O(2,1)}
\end{equation}%
and the action of the vector multiplet of creation operators on these states
generates complicated functions that do not even approximate the first
excited levels. \-\-Apparently, the vector operator multiplet belongs only
to the $s=0$ vector representations of the symmetry groups, and not to the
infinite-dimensional $s=1/2$ representations. It may be possible to
construct an appropriate ladder representation of creation/annihilation
operators though a multipole expansion of the Hamiltonian, corresponding to
an infinite summation of the associated Legendre functions. This will be
discussed in a subsequent paper.

A working mode number representation for the $s=1/2$ representations of
O(2,1) is required to clarify the question of ghost states for the
relativistic harmonic oscillator. As seen in (\ref{ghost}) excited timelike
modes of the Feynman, Kislinger, and Ravndal wavefunctions may have negative
norm, which were handled in \cite{FKR} by applying the covariant condition%
\begin{equation}
\left( p\cdot a\right) \psi =0  \label{GB}
\end{equation}%
which forces $\psi $ into the ground state along the momentum $p$,
suppressing timelike excitations. Although this approach is comparable to
Gupta-Bleuler quantization \cite{IZ} of the electromagnetic field, where (%
\ref{GB}) expresses the Lorentz gauge condition as an operator equation and
eliminates negative norm states along the field's lightlike momentum, there
is no applicable gauge condition for the general relativistic oscillator
that justifies this procedure.

Interestingly, the problem of negative normed states could have been
inadvertently overlooked without reference to the creation/annihilation
operators, because without sufficient attention to the properties of the
states as representations of the Lorentz group, we might neglect to include
the metric in our calculations. For example, the first excited timelike mode
is%
\begin{equation}
\psi _{1}^{0}=\bar{a}^{0}\psi _{0}=\frac{1}{\sqrt{2}}\left( x^{0}-\partial
^{0}\right) A_{0}e^{-\rho ^{2}/2}=A_{0}\sqrt{2}te^{-\rho ^{2}/2}
\end{equation}%
and we may follow the method of (\ref{ghost}) to calculate%
\begin{eqnarray}
\int dt~d^{2}x~\left\vert \psi _{1}^{0}\right\vert ^{2} &=&\int
dt~d^{2}x~\left\vert \overline{a}^{0}\psi _{0}\right\vert ^{2}=\frac{1}{2}%
\int dt~d^{2}x~\left( \left( x^{0}-\partial ^{0}\right) \psi _{0}\right)
^{\dagger }\left( \left( x^{0}-\partial ^{0}\right) \psi _{0}\right)  \notag
\\
&=&\int dt~d^{2}x~\frac{1}{2}\psi _{0}^{\ast }\left( x^{0}+\partial
^{0}\right) \left( x^{0}-\partial ^{0}\right) \psi _{0}  \notag \\
&=&\int dt~d^{2}x~\frac{1}{2}\psi _{0}^{\ast }\left[ \left( x^{0}-\partial
^{0}\right) \left( x^{0}+\partial ^{0}\right) +2\eta ^{00}\right] \psi _{0} 
\notag \\
&=&-\int dt~d^{2}x~\psi _{0}^{\ast }\psi _{0}<0  \label{norm}
\end{eqnarray}%
where we use 
\begin{equation}
\left( x^{0}-\partial _{0}\right) \psi _{0}=0
\end{equation}%
and assume some regularization for the ground state normalization. However,
neglecting to include the metric in the formulation of the norm, we might be
tempted to calculate%
\begin{equation}
\int dt~d^{2}x~\left\vert \psi _{1}^{0}\right\vert ^{2}=\int
dt~d^{2}x~\left\vert A_{0}\sqrt{2}te^{-\rho ^{2}/2}\right\vert ^{2}=2\int
dt~d^{2}x~t^{2}\psi _{0}^{\ast }\psi _{0}>0  \label{norm2}
\end{equation}%
which contradicts (\ref{norm}). Given the role played by the metric in (\ref%
{norm}), it seems that the proper formulation of the norm requires that we
respect the tensor properties of each excited state and not inadvertantly
treat the states as scalar entities. Thus, the norm (\ref{norm2}) should be
restated as%
\begin{equation}
\int dt~d^{2}x~\left\Vert \psi _{nlm}\right\Vert ^{2}=\int dt~d^{2}x~\eta
_{mm}\left( \psi _{nlm}\right) ^{\dagger }\psi _{nlm}
\end{equation}%
where $\eta _{mm}$ represents the metric in the relevant tensor
representation. In the absence of a number representation for the
relativistic oscillator, the straightforward calculation in (\ref{ghost})
cannot be performed to check that no ghosts appear in this formulation. We
may argue that since the wavefunctions (\ref{solu-5}) are not separable into
Cartesian modes, all polar modes mix space and time within the spacelike
sector, and so there should be no timelike excitations as such in the
relativistic oscillator. Moreover, given the infinite dimensional multiplets
of states, there is no particular state that is naturally assigned a
negative metric. These claims will receive more detailed treatment in a
subsequent paper.


\begin{thebibliography}{99}
\bibitem{Kastrup} Hans Kastrup, \textit{Annalen Phys. }\textbf{16} 439
(2007).

\bibitem{Bars} Itzhak Bars, \textit{Phys. Rev. D} \textbf{79}, 045009 (2009).

\bibitem{Bars-comment} I am grateful to Itzhak Bars for this observation.
See also, J. Schwinger, \textquotedblleft On Angular
Momentum\textquotedblright , in \textit{Quantum Theory of Angular Momentum},
edited by L. C. Biedenharn and H. Van Dam, (Academic Press, New York, 1965),
page 229 (available from
http://www.osti.gov/accomplishments/documents/fullText/ACC0111.pdf), and F.
M. Mej\'{\i}a and V. Pleitez, \textit{Rev. Bras. Ens. Fis.} \textbf{24} 41
(2002).

\bibitem{bound} L.\ P.\ Horwitz and R.\ Arshansky, \textit{J.\ Math.\ Phys.} 
\textbf{30} 66 (1989); \newline
\textit{J.\ Math.\ Phys.} \textbf{30} 380 (1989).

\bibitem{H-P} L.\ P.\ Horwitz and C.\ Piron, \textit{Helv.\ Phys.\ Acta} 
\textbf{48} 316 (1973).

\bibitem{talk} Martin Land, \textit{On timelike excitations in the
relativistic harmonic oscillator}, lecture slides available from
http//www.hadassah.ac.il/CS/staff/martin/iard2008.pdf.

\bibitem{K-N} Y. S. Kim and Marilyn E. Noz, \textit{Phys. Rev.} \textit{D} 
\textbf{8} 3521 (1973).

\bibitem{FKR} R. P. Feynman, M. Kislinger, and F. Ravndal, \textit{Phys. Rev.%
} \textit{D} \textbf{3} 2706 (1971).

\bibitem{Gradshteyn and Ryzhik} I. S. Gradshteyn and I. M. Ryzhik, \textit{%
Table of Integrals, Series, and Products}, Academic Press, London, 1994.

\bibitem{IZ} C. Itzykson J.-B. Zuber, \textit{Quantum Field Theory},
McGraw-Hill, New York, 1980.
\end{thebibliography}
\end{document}